\let\userbibliographystyle\bibliographystyle
\renewcommand{\bibliographystyle}[1]{}
\documentclass[pdflatex,sn-mathphys-num]{sn-jnl}
\let\bibliographystyle\userbibliographystyle

\usepackage{graphicx}
\usepackage{multirow}
\usepackage{amsmath,amssymb,amsfonts}
\usepackage{amsthm}
\usepackage{mathrsfs}
\usepackage[title]{appendix}
\usepackage{xcolor}
\usepackage{textcomp}
\usepackage{manyfoot}
\usepackage{booktabs}
\usepackage{algorithm}
\usepackage{algorithmicx}
\usepackage{algpseudocode}
\usepackage{listings}
\usepackage{comment}
\usepackage{color}
\usepackage{overpic}
\usepackage{epsfig}
\usepackage{epstopdf}
\usepackage{colortbl}
\usepackage{subfigure}
\usepackage{rotating}
\usepackage{bm}
\usepackage{dcolumn}
\usepackage{makecell}
\usepackage{xspace}
\usepackage{xparse}
\usepackage{tabularx}
\usepackage{placeins}
\usepackage{hyperref}

\providecommand{\doi}[1]{}
\renewcommand{\doi}[1]{\href{https://doi.org/#1}{https://doi.org/#1}}
\hypersetup{
  colorlinks=true,
  linkcolor=blue!55!black,
  citecolor=blue!65!black,
  urlcolor=blue!55!black,
  pdfborder={0 0 0}
}
\NewDocumentCommand{\STCFMC}{o m}{%
  \IfNoValueTF{#1}{\textsc{#2}\xspace}{\textsc{#2}\,#1\xspace}}
\NewDocumentCommand{\PYTHIA}{o}{\STCFMC[#1]{Pythia}}

\theoremstyle{thmstyleone}

\theoremstyle{thmstyletwo}

\theoremstyle{thmstylethree}

\begin{document}

\title{Entanglement-sensitive observables in $e^+e^- \to \tau^+\tau^-$ at STCF: a detector-level feasibility study}

\author[1]{\fnm{Chentao} \sur{Bao}}
\author[2,3]{\fnm{Xi} \sur{Tao}}
\author*[1]{\fnm{Hai} \sur{Chen}}\email{haichen@zju.edu.cn}
\author*[2,3]{\fnm{Lailin} \sur{Xu}}\email{lailinxu@ustc.edu.cn}
\author[2,3]{\fnm{Xiaorong} \sur{Zhou}}
\author*[2,3]{\fnm{Mingyi} \sur{Liu}}\email{nh196245@ustc.edu.cn}

\affil[1]{\orgdiv{School of Physics}, \orgname{Zhejiang University}, \orgaddress{\city{Hangzhou}, \postcode{310058}, \country{People's Republic of China}}}
\affil[2]{\orgdiv{Department of Modern Physics}, \orgname{University of Science and Technology of China}, \orgaddress{\city{Hefei}, \postcode{230026}, \country{People's Republic of China}}}
\affil[3]{\orgdiv{State Key Laboratory of Particle Detection and Electronics}, \orgaddress{\city{Hefei}, \postcode{230026}, \country{People's Republic of China}}}

\abstract{
Quantum entanglement in $\tau^+\tau^-$ production provides a direct probe of non-classical spin correlations in a relativistic quantum system, with the $\tau$ decay products serving as spin analyzers. We investigate the detector-level feasibility of such measurements in $e^+e^-\to\tau^+\tau^-$ at the Super Tau-Charm Facility (STCF) at $\sqrt{s}=7$~GeV, using the $\rho\rho$ and $\pi\rho$ decay channels. Signal and dominant background processes are simulated with the full STCF detector and reconstruction framework. Dedicated reconstruction and event-selection procedures yield overall signal efficiencies of approximately $4.5\%$ and $4.8\%$, with corresponding purities of about $87.6\%$ and $93.2\%$ for the $\rho\rho$ and $\pi\rho$ channels, respectively. From the reconstructed decay kinematics, the $\tau^+\tau^-$ spin-density matrix is inferred and the concurrence $C[\rho]$ together with the Bell-sensitive quantity $m_{12}[C]$ are evaluated in a selected fiducial region. For the benchmark requirement $|\cos\theta|<0.2$ and an integrated luminosity of $1~\mathrm{ab}^{-1}$, the projected relative precision on $C[\rho]$ is about $2.2\%$ in both channels, while that on $m_{12}[C]$ is approximately $1.4\%$ for $\rho\rho$ and $1.5\%$ for $\pi\rho$, including statistical and detector-resolution uncertainties. The reconstructed $m_{12}[C]$ shows a larger separation from its reference threshold in the $\rho\rho$ channel, making this topology particularly sensitive to Bell-type spin correlations. These results demonstrate the detector-level feasibility of probing quantum entanglement and non-classical spin correlations in hadronic $\tau$ decays at STCF.
}

\keywords{$\tau$-lepton physics, quantum entanglement, spin correlations, Bell-sensitive observables, Super Tau-Charm Facility}

\maketitle

\section{Introduction}
\label{sec:intro}


Quantum entanglement is a defining feature of quantum mechanics and provides a direct probe of correlations that cannot be described by separable classical states. While experimental studies of entanglement and Bell nonlocality have historically been performed mainly with photons, atoms, and other low-energy quantum systems, recent developments have extended these concepts to high-energy particle collisions~\cite{clauser:1969,Horodecki:2009zz,Barr:2024djo}. At colliders, the spin degrees of freedom of pair-produced particles provide a natural bipartite quantum system, and their spin-density matrix can be inferred from angular correlations among the decay products. This has motivated a growing program of quantum-information measurements and feasibility studies involving top quarks, hyperons, and leptons~\cite{Fabbrichesi:2023,Atlas:2024,ZhangNeutrino:2026}.


Lepton colliders provide a particularly clean environment for extending quantum-entanglement studies to elementary-particle systems, owing to their well-defined initial states and the possibility of reconstructing spin information from correlated decay products. Recent theoretical and experimental developments have demonstrated the potential of this approach in several processes. At BESIII, entanglement and Bell nonlocality have been investigated in hyperon--antihyperon production~\cite{PhysRevD.110.054012}, and a subsequent measurement using entangled $\Lambda\bar{\Lambda}$ pairs provided an experimental test of local-realistic descriptions~\cite{BESIII:2025entanglement}. These studies illustrate that quantum-information observables can be accessed in realistic $e^+e^-$ collision environments, including the effects of detector acceptance, reconstruction, and backgrounds.


Among the processes accessible at lepton colliders, $e^+e^-\to\tau^+\tau^-$ provides a particularly attractive system for studying spin entanglement. The production mechanism is theoretically well controlled, while the parity-violating weak decays of the $\tau$ leptons retain information on their spin orientations through the kinematics of the visible decay products. The one-prong hadronic modes $\tau^\pm\to\pi^\pm\nu_\tau$ and $\tau^\pm\to\rho^\pm\nu_\tau\to\pi^\pm\pi^0\nu_\tau$ are especially useful because of their sizable branching fractions, simple visible topologies, and strong spin-analyzing power~\cite{Jadach:1990mz}. A dedicated feasibility study for Belle~II considered $e^+e^-\to\tau^+\tau^-$ at $\sqrt{s}=10.579$~GeV using several hadronic $\tau$ decay modes and showed that the spin-density matrix, entanglement, and Bell-sensitive observables can in principle be reconstructed with high sensitivity when experimental resolutions are taken into account~\cite{PhysRevD.109.032005}. More recently, the prospects for entanglement and Bell-nonlocality studies in $\tau^+\tau^-$ production near and above threshold at BEPC have also been investigated, emphasizing the importance of detector effects, acceptance, and systematic control for a realistic measurement~\cite{HanLowSu:2025}.


The Super Tau-Charm Facility (STCF) would provide a particularly favorable and distinctive environment for extending these studies. It is designed to operate over the center-of-mass energy range $2$--$7$~GeV with high luminosity, enabling large samples of $\tau^+\tau^-$ events to be collected in the intermediate-energy region between production threshold and the $B$-factory scale~\cite{achasov2024stcf}. Its tunable collision energy provides an opportunity to investigate the energy dependence of the $\tau$-pair spin-correlation structure, while the expected tracking, particle-identification, and calorimeter performance enables detailed reconstruction of hadronic $\tau$ decays. Within the STCF energy range, the Standard Model predicts increasingly pronounced entanglement-sensitive spin correlations toward higher center-of-mass energies, particularly for centrally produced $\tau$ pairs, motivating the choice of $\sqrt{s}=7$~GeV for the present study. Together, the high event yield, tunable collision energy, and dedicated detector capabilities establish STCF as a distinct experimental setting for $\tau$-pair quantum-correlation studies, extending the investigation of these observables into an energy regime complementary to those explored at BEPC and the higher-energy Belle~II experiment.

In this work, we perform a detector-level feasibility study of $e^+e^-\to\gamma^\ast\to\tau^+\tau^-$ at STCF using the $\pi^\pm\rho^\mp$ and $\rho^+\rho^-$ final states, where $\pi$ and $\rho$ denote the $\tau^\pm\to\pi^\pm\nu_\tau$ and $\tau^\pm\to\rho^\pm\nu_\tau\to\pi^\pm\pi^0\nu_\tau$ decay modes, respectively. The two channels provide complementary experimental
topologies: the $\pi\rho$ channel has a simpler neutral final state,
whereas the $\rho\rho$ channel provides polarimetric information from
both $\rho$ decays. The complementary $\pi^+\pi^-$ channel has been investigated separately in an earlier STCF study~\cite{Li:2026wxi}. Signal and dominant background processes are simulated with the full STCF detector and reconstruction framework, and the analysis includes charged-track and photon selection, $\pi^0$ and $\rho$ reconstruction, constrained kinematic fitting, multivariate background suppression, and reconstruction of the $\tau$ flight directions. From the reconstructed decay kinematics, we infer the $\tau^+\tau^-$ spin-density matrix and evaluate the concurrence $C[\rho]$ as an entanglement witness and $m_{12}[C]$ as a Bell-sensitive witness of the reconstructed spin-correlation matrix. The aim is not to perform a loophole-free Bell test or to extrapolate the observables to the full phase space, but to quantify their achievable detector-level sensitivity in experimentally accessible fiducial regions, with statistical uncertainties evaluated by bootstrap resampling and detector-resolution effects propagated using a covariance-based smearing procedure.

\section{Spin-correlation formalism and entanglement-sensitive observables}
\label{sec:theory}

The spin information carried by the $\tau^+\tau^-$ system is inferred from the kinematics of its decay products through a reconstruction of the two-particle spin-density matrix (SDM). In the present analysis, the $\pi\nu_\tau$ and $\rho\nu_\tau$ decay modes provide the corresponding polarimetric information, allowing the single-spin polarization coefficients and spin-correlation matrix to be estimated event by event. The reconstructed SDM then provides the basis for evaluating the concurrence $C[\rho]$ and the Bell-sensitive quantity $m_{12}[C]$ discussed below.

\subsection{Spin-density matrix and observable coefficients}
\label{subsec:sdm}

In the two-particle spin space, the most general SDM of the $\tau^+\tau^-$ system can be written as
\begin{equation}
\begin{aligned}
\rho_{\tau^+\tau^-}=\frac{1}{4}\Big[&\mathbf{1}\otimes\mathbf{1}
+\sum_i B_i^+\,\sigma_i\otimes\mathbf{1}+\sum_i B_i^-\,\mathbf{1}\otimes\sigma_i
+\sum_{i,j} C_{ij}\,\sigma_i\otimes\sigma_j\Big].
\end{aligned}
\label{eq:sdm_general}
\end{equation}
where $\sigma_i$ ($i=1,2,3$) are the Pauli matrices, $B_i^\pm$ denotes the polarization components of the $\tau^\pm$ leptons, and $C_{ij}$ are the elements of the spin-correlation matrix. Together, the vectors $B_i^\pm$ and the matrix $C_{ij}$ specify the spin structure of the reconstructed two-particle state. The entanglement-sensitive observables considered in this work are determined primarily by the correlation matrix $C$.

The coefficients of Eq.~\eqref{eq:sdm_general} can be inferred from the polarimetric vectors $\vec{h}^{\,\pm}$ associated with the two $\tau$ decays. For ideal acceptance and integration over the full angular phase space, the single-spin and spin-correlation coefficients can be written as~\cite{Barr:2024djo}
\begin{equation}
\begin{aligned}
B_i^\pm=\frac{3}{\kappa_\pm\sigma}
\int d\Omega_\pm\,
\frac{d\sigma}{d\Omega_\pm}
\left(\vec{h}^{\,\pm}\!\cdot\hat{e}_i\right)=3\langle h_i^\pm\rangle.
\end{aligned}
\label{eq:Bi_def}
\end{equation}
\begin{equation}
\begin{aligned}
C_{ij}=\frac{9}{\kappa_+\kappa_-\sigma}
\int d\Omega_+\,d\Omega_-\,
\frac{d\sigma}{d\Omega_+\,d\Omega_-}\times
\left(\vec{h}^{\,+}\!\cdot\hat{e}_i\right)
\left(\vec{h}^{\,-}\!\cdot\hat{e}_j\right)=9\langle h_i^+ h_j^-\rangle.
\end{aligned}
\label{eq:Cij_def}
\end{equation}
where $\kappa_\pm$ denote the spin-analyzing powers and $\hat e_i$ are the unit vectors of the chosen spin basis. With the polarimetric-vector definitions adopted in this work, the analyzing power is absorbed into $\vec h^{\,\pm}$, corresponding to $\kappa_\pm=1$ for the $\pi\nu_\tau$ and $\rho\nu_\tau$ modes considered below.


The moment relations in Eqs.~\ref{eq:Bi_def} and~\ref{eq:Cij_def} correspond to the ideal case of full angular coverage and unbiased phase-space integration. In the present study, the spin observables are evaluated within a specified fiducial phase-space region, thereby characterizing the $\tau^+\tau^-$ spin correlations of the selected event ensemble. At detector level, the quantities $h_i^\pm$ and $h_i^+h_j^-$ are evaluated from reconstructed events after the full event selection, so their sample averages also incorporate the effects of detector acceptance and reconstruction efficiency. The resulting $B_i^\pm$ and $C_{ij}$ are therefore treated as reconstructed fiducial estimators of the spin-density matrix, and the derived quantities $C[\rho]$ and $m_{12}[C]$ are interpreted consistently within the same detector-level fiducial definition. This formulation is appropriate for the present feasibility study, whose aim is to quantify the achievable sensitivity to entanglement-sensitive spin correlations under realistic reconstruction and selection conditions. 

\subsection{Coordinate basis and decay polarimetry}

The spin-correlation coefficients are evaluated in the orthonormal basis $\{\hat n,\hat r,\hat k\}$ defined in the $\tau^+\tau^-$ center-of-mass frame by
\begin{equation}
\hat{n}
=
\frac{\hat{p}\times\hat{k}}{\sin\theta},
\qquad
\hat{r}
=
\frac{\hat{p}-\cos\theta\,\hat{k}}{\sin\theta},
\label{eq:basis_def}
\end{equation}
where $\hat k$ is chosen along the $\tau^-$ flight direction, $\hat p$ denotes the electron-beam direction, and $\theta$ is the production polar angle of the $\tau^-$. In the following, the indices $i=1,2,3$ correspond to the $\hat n$, $\hat r$, and $\hat k$ directions, respectively.

The polarimetric vector depends on the $\tau$ decay mode. For the two-body decay $\tau^\pm\to\pi^\pm\nu_\tau$, the visible pion provides a maximal spin analyzer, and the polarimetric vector is given by
\begin{equation}
\vec{h}=-\hat{n}_\pi,
\label{eq:h_pi_mode}
\end{equation}
where $\hat n_\pi$ denotes the pion direction in the corresponding $\tau$ rest frame. This expression is used for the $\pi\nu_\tau$ side of the $\pi\rho$ channel.

For $\tau^\pm\to\rho^\pm\nu_\tau\to\pi^\pm\pi^0\nu_\tau$, both the orientation of the visible $\rho$ system and its internal decay kinematics carry spin information. Following Ref.~\cite{Jadach:1990mz}, we first define the spatial polarimetric analyzer
\begin{equation}
\vec{H}
=
-\left[
2(q^\mu N_\mu)\,\vec{q}
-
(q^\mu q_\mu)\,\vec{N}
\right],
\label{eq:rho_polarimeter_raw}
\end{equation}
where $q^\mu=q^\mu_{\pi^\pm}-q^\mu_{\pi^0}$, and $N^\mu$ is the neutrino four-momentum. The polarimetric vector used in the present analysis is the corresponding unit vector,
\begin{equation}
\vec{h}
=
\frac{\vec{H}}{|\vec{H}|}.
\label{eq:rho_polarimeter}
\end{equation}
Thus, only the direction of the polarimetric analyzer is retained, and $|\vec{h}|=1$ event by event. This normalized polarimeter is used for both $\tau$ decays in the $\rho\rho$ channel and for the $\rho\nu_\tau$ side of the $\pi\rho$ channel.

The construction of these polarimetric vectors is directly matched to the reconstruction procedure described in Sec.~\ref{sec:selection}. After the visible final state has been assigned to the two $\tau$ decay chains and the $\tau$ flight directions have been reconstructed, the decay products are boosted to the corresponding $\tau$ rest frames. Eq.~\eqref{eq:h_pi_mode} is then used for a $\pi\nu_\tau$ decay and Eq.~\eqref{eq:rho_polarimeter} for a $\rho\nu_\tau$ decay. The resulting event-level components $h_i^\pm$ and products $h_i^+h_j^-$ provide the estimators entering the reconstructed polarization and spin-correlation coefficients.

\subsection{Derived entanglement-sensitive observables}
\label{subsec:entangle_sensitive}

Two quantities derived from the reconstructed SDM are used to characterize non-classical spin correlations in this study: the concurrence $C[\rho]$ and the quantity $m_{12}[C]$ associated with the Clauser-Horne-Shimony-Holt (CHSH) criterion~\cite{clauser:1969}.

The concurrence is a standard measure of entanglement for a two-qubit state. A separable state can be expressed as
\begin{equation}
\begin{aligned}
\rho_{\mathrm{sep}}
&=\sum_{a,b}p_{ab}\,
\rho_{\tau^+}^{(a)}\otimes\rho_{\tau^-}^{(b)},\qquad p_{ab}>0,\qquad \sum_{a,b}p_{ab}=1.
\end{aligned}
\label{eq:sep_state}
\end{equation}
where $\rho_{\tau^\pm}^{(a)}$ denote the density matrices of the individual $\tau^\pm$ spin states, and $p_{ab}$ are the non-negative classical mixing weights of the corresponding product states. The concurrence is defined as~\cite{Bennett:1996,Wootters:1998}
\begin{equation}
\begin{aligned}
C[\rho]&=\max\{0,\lambda_1-\lambda_2-\lambda_3-\lambda_4\},\hspace{0.5cm}0\le C[\rho]\le1.
\end{aligned}
\label{eq:concurrence_def}
\end{equation}
where $\lambda_i$ are the eigenvalues, in decreasing order, of
\begin{equation}
\begin{aligned}
R=\sqrt{\sqrt{\rho}\,\tilde{\rho}\,\sqrt{\rho}},\hspace{0.5cm}
\tilde{\rho}=(\sigma_2\otimes\sigma_2)\,\rho^\ast\,
(\sigma_2\otimes\sigma_2).
\end{aligned}
\label{eq:R_def}
\end{equation}
For a physical two-qubit density matrix, $C[\rho]=0$ for separable states, whereas $C[\rho]>0$ indicates entanglement; $C[\rho]=1$ corresponds to a maximally entangled pure state.

The second quantity, $m_{12}[C]$, characterizes the maximal CHSH correlation associated with the reconstructed spin-correlation matrix. For a two-qubit state, the CHSH combination may be written as~\cite{Horodecki:1995}
\begin{equation}
\begin{aligned}
\big|&\hat{n}_1\cdot C\cdot(\hat{n}_2-\hat{n}_4)+\hat{n}_3\cdot C\cdot(\hat{n}_2+\hat{n}_4)\big|\le2.
\end{aligned}
\label{eq:CHSH_form}
\end{equation}
where $\hat{n}_i$ are unit vectors specifying the measurement axes. Defining
\begin{equation}
M=C^T C,
\label{eq:M_def}
\end{equation}
and denoting by $m_1$ and $m_2$ the two largest eigenvalues of $M$, one obtains
\begin{equation}
m_{12}[C]=m_1+m_2.
\label{eq:m12_def}
\end{equation}
For a physical two-qubit state, $m_{12}[C]>1$ constitutes the Horodecki necessary and sufficient condition for violation of the CHSH inequality~\cite{Horodecki:1995}.

The interpretation of $m_{12}[C]$ in the present collider analysis requires an important qualification. The spin information is inferred from $\tau$ decay products rather than obtained from freely chosen spin measurements, and the effective analysis axes are defined from the event kinematics. The procedure therefore does not constitute a loophole-free Bell test with independently selectable measurement settings. We consequently use $m_{12}[C]$ as a Bell-sensitive observable derived from the reconstructed spin-correlation matrix, while $C[\rho]$ quantifies the entanglement of the reconstructed two-spin state.

\begin{figure}[htbp]
  \centering
  \includegraphics[width=0.48\textwidth]{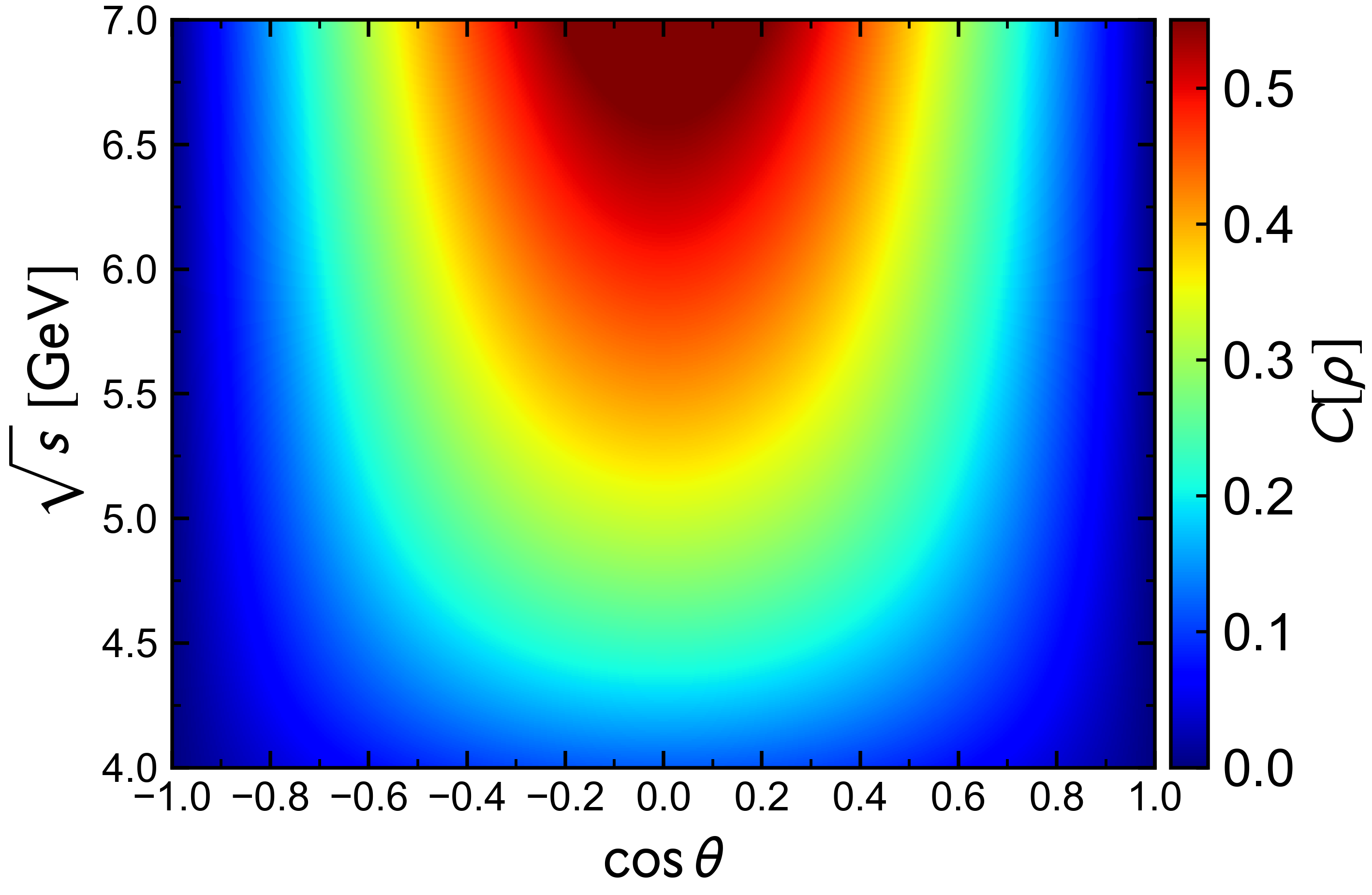}~~
  \includegraphics[width=0.48\textwidth]{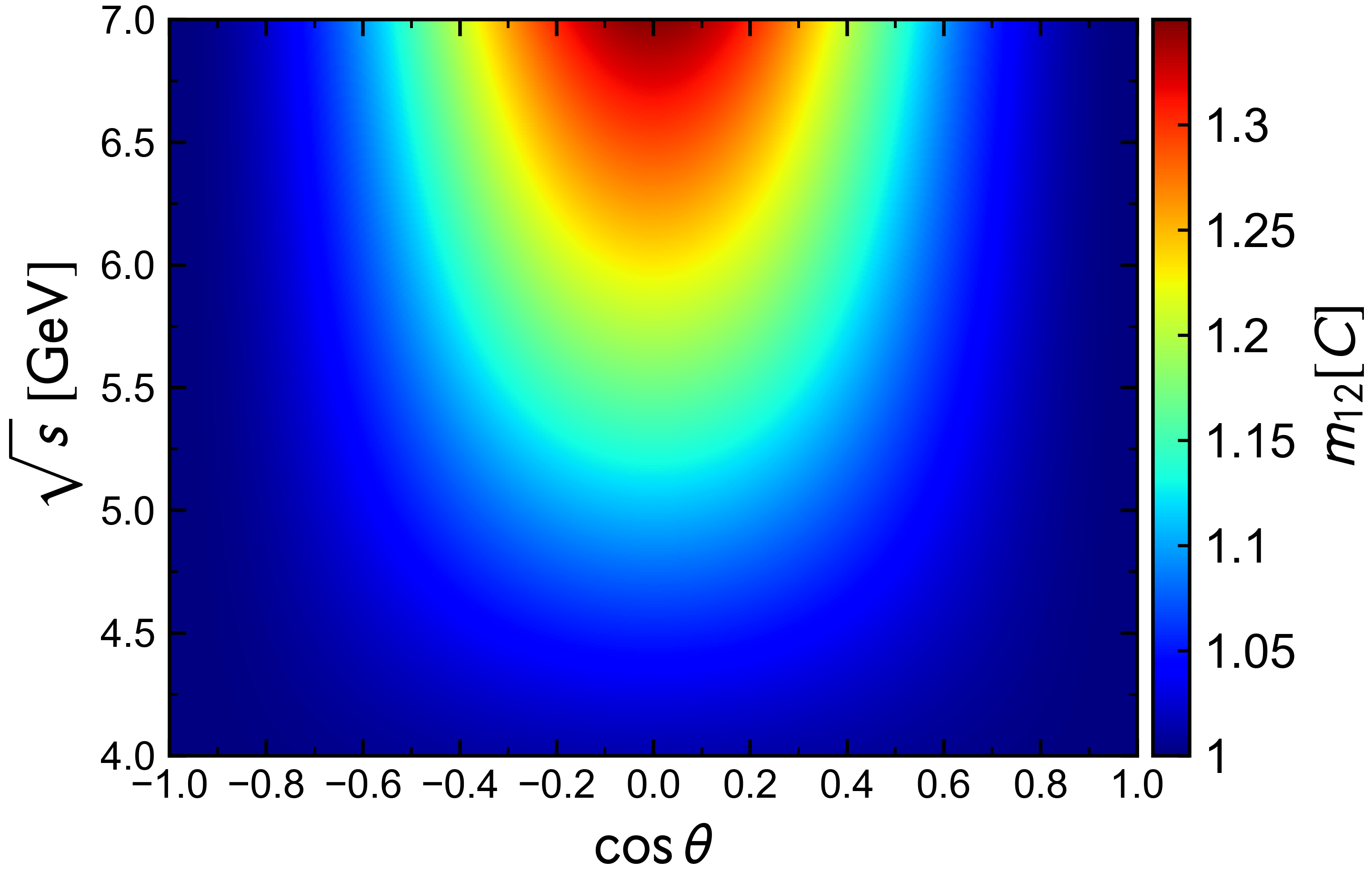}
    \caption{Standard Model expectations for the concurrence $C[\rho]$ (left) and the Bell-sensitive quantity $m_{12}[C]$ (right) as functions of the center-of-mass energy $\sqrt{s}$ and the $\tau^-$ production angle $\theta$ in the $\tau^+\tau^-$ center-of-mass frame, based on the analytical calculation of Ref.~\cite{PhysRevD.109.032005}.}
  \label{fig:QEwitness}
\end{figure}

Figure~\ref{fig:QEwitness} shows the Standard Model predictions for $C[\rho]$ and $m_{12}[C]$ as functions of the center-of-mass energy and the $\tau$ production angle $\theta$, defined as the polar angle of the $\tau^-$ momentum with respect to the electron beam direction in the $\tau^+\tau^-$ center-of-mass frame~\cite{PhysRevD.109.032005}. Within the STCF energy range, both observables become more pronounced toward higher center-of-mass energies and for centrally produced $\tau$ pairs. This behavior motivates the choice of $\sqrt{s}=7$~GeV for the present detector-level study and the subsequent optimization of the fiducial requirement on $|\cos\theta|$.

The observable sensitivity at detector level is determined by the interplay between the underlying spin-correlation structure and the experimental reconstruction efficiency and resolution. The event selection and $\tau$-direction reconstruction described in Sec.~\ref{sec:selection} are therefore optimized to retain the regions of phase space with favorable polarimetric sensitivity while maintaining sufficient event yield. The resulting fiducial dependence of $C[\rho]$ and $m_{12}[C]$ is examined quantitatively in Sec.~\ref{sec:result}.

\section{The STCF detector and Monte Carlo samples}
\label{sec:detector_mc}

The Super Tau-Charm Facility (STCF) is a proposed high-luminosity $e^+e^-$ collider designed to operate at center-of-mass energies between 2 and 7~GeV, with a peak luminosity of $0.5\times10^{35}~\mathrm{cm}^{-2}\mathrm{s}^{-1}$ at $\sqrt{s}=4.0$~GeV~\cite{achasov2024stcf}. Its baseline detector concept comprises an inner tracking system based on either $\mu$-RWELL gaseous detectors or MAPS silicon sensors, a main drift chamber (MDC), a particle-identification system, an electromagnetic calorimeter (EMC), a 1~T superconducting solenoid, and a muon detector. For the present analysis, the most relevant detector capabilities are charged-track reconstruction, photon and $\pi^0$ reconstruction, particle identification, and the trajectory information used in the reconstruction of the $\tau$ flight direction.

Detector simulation, digitization, and event reconstruction are performed within the STCF offline software framework OSCAR~\cite{huang:2023}. Signal and background Monte Carlo samples are generated at the benchmark center-of-mass energy $\sqrt{s}=7.0$~GeV. The hard process $e^+e^-\to\tau^+\tau^-$ is generated at leading order with \textsc{MadGraph5\_aMC@NLO}~\cite{Alwall:2014}, while initial-state radiation is modeled through the interface to \textsc{Pythia}~8.306~\cite{Bierlich:2022}. All generated events are subsequently processed through the full detector simulation and standard OSCAR reconstruction chain.

Two signal topologies are considered:
\[
\tau^+\tau^- \to (\pi^\pm\nu_\tau)(\rho^\mp\nu_\tau), \qquad
\tau^+\tau^- \to (\rho^+\nu_\tau)(\rho^-\nu_\tau),
\]
with $\rho^\pm\to\pi^\pm\pi^0$ and $\pi^0\to\gamma\gamma$. In the following, these are referred to as the $\pi\rho$ and $\rho\rho$ channels, respectively. Both final states contain two charged pions and two undetected neutrinos, while the $\pi\rho$ ($\rho\rho$) topology contains one (two) reconstructed $\pi^0$ mesons and therefore two (four) signal photons. Additional neutral EMC clusters may arise from initial-state radiation or detector effects.

The dominant $\tau$ decay modes are simulated with the Decay Chain functionality of \textsc{MadGraph5\_aMC@NLO}, together with the extended $\tau$-decay implementation provided by the \texttt{taudecay} UFO package~\cite{Hagiwara:2012taudecay}, while minor decay modes are generated with \textsc{Pythia}~8.306. The branching fractions and generator assignments for the principal decay categories are summarized in Table~\ref{tab:tau_decay_tools}. For events in which both $\tau$ decays are handled within the Decay Chain framework, the spin correlations between the two $\tau$ leptons and their decay products are retained explicitly. The remaining minor decay modes generated with \textsc{Pythia} are included as background contributions in the event-selection study.

\begin{table}[tb]
\centering
\caption{Branching fractions~\cite{PDG:2024} and simulation tools used for the principal $\tau$ decay modes included in the Monte Carlo samples. Minor decay modes not listed individually are generated with \textsc{Pythia}~8.306.}
\label{tab:tau_decay_tools}
\begin{tabular}{lcc}
\hline
$\tau$ decay mode & Branching fraction & Simulation tool \\
\hline
$\pi\nu_\tau$ & $10.8\%$ & \multirow{6}{*}{\textsc{Decay Chain}} \\
$\pi\pi^0\nu_\tau$ & $25.5\%$ & \\
$\pi\pi^0\pi^0\nu_\tau$ & $9.3\%$ & \\
$\pi\pi\pi\nu_\tau$ & $9.3\%$ & \\
$e\nu_e\nu_\tau$ & $17.8\%$ & \\
$\mu\nu_\mu\nu_\tau$ & $17.4\%$ & \\
\hline
Minor modes & $9.9\%$ & {\sc Pythia}~8.306 \\
\hline
\end{tabular}
\end{table}

The dominant backgrounds to the $\pi\rho$ and $\rho\rho$ decay channels arise from other $\tau^+\tau^-$ decay modes, particularly hadronic channels that can reproduce the selected charged-track and photon multiplicities through particle misidentification, missing particles, or combinatorial assignment. Representative contributions include $(\rho\nu)(\pi\pi^0\pi^0\nu)$, $(\rho\nu)(K^\ast\nu)$, and other multipion final states. These backgrounds are included explicitly in the simulation. Non-resonant hadronic production, $e^+e^-\to q\bar q$, is also simulated using \textsc{MadGraph5\_aMC@NLO} for the hard process and \textsc{Pythia}~8.306 for hadronization.

Processes such as $e^+e^-\to\mu^+\mu^-$, Bhabha scattering, $e^+e^-\to\gamma\gamma$, and two-photon production are not included in the nominal Monte Carlo samples. Their final-state topologies differ substantially from the selected hadronic $\tau$ signatures and are expected to be strongly reduced by the charged-particle multiplicity, pion-identification, photon, missing-momentum, and kinematic-fit requirements. Since these contributions are not explicitly evaluated in the present study, their residual impact is not included in the quoted background estimates. Beam-induced and other machine-related backgrounds are likewise beyond the scope of the present detector-level feasibility study.

At $\sqrt{s}=7$~GeV, the anticipated STCF luminosity is expected to provide large samples of $\tau^+\tau^-$ events, including sizeable $\pi\rho$ and $\rho\rho$ subsamples. These two channels therefore provide experimentally relevant benchmarks for assessing the reconstruction of spin correlations and entanglement-sensitive observables with the full STCF detector response. The event-selection and reconstruction strategies developed for the two topologies are described in Sec.~\ref{sec:selection}.

\section{Event reconstruction and selection}
\label{sec:selection}

\begin{figure}[htbp]
  \centering
  \includegraphics[width=0.76\textwidth]{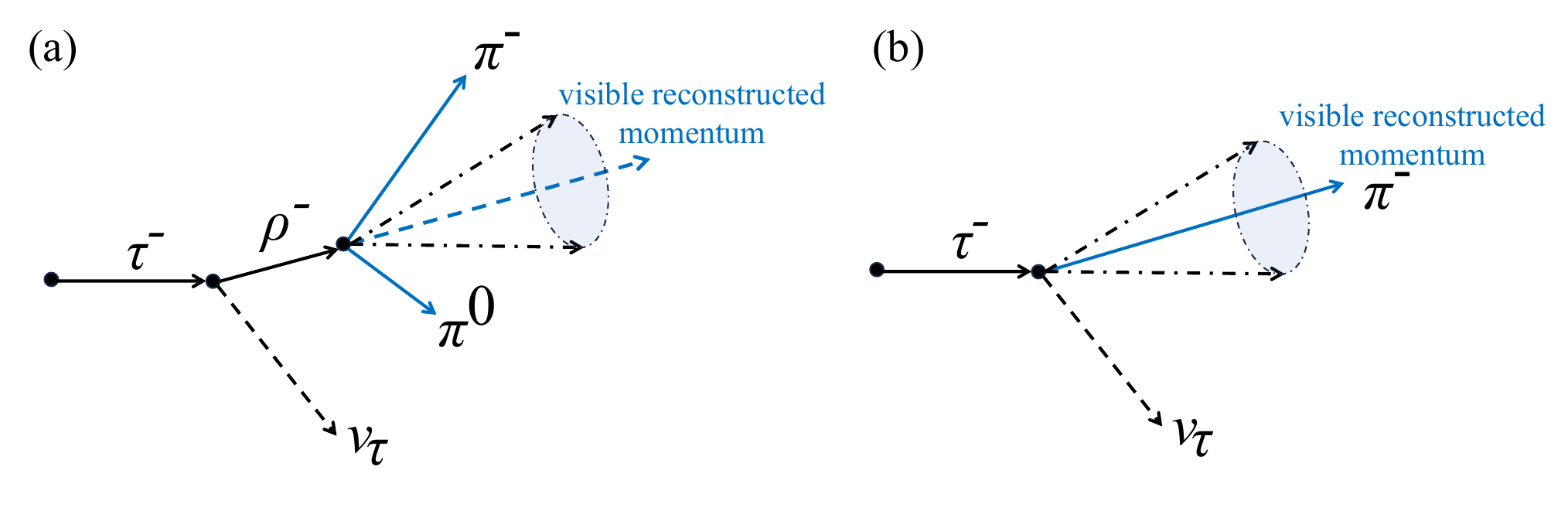}
    \caption{Schematic kinematic construction for $\tau^-\to\rho^-\nu_\tau$, $\rho^-\to\pi^-\pi^0$ (left), and $\tau^-\to\pi^-\nu_\tau$ (right). In each case, the undetected neutrino constrains the parent $\tau^-$ direction to lie on a cone around the visible hadronic momentum. For the $\rho\nu_\tau$ decay, the cone axis is the reconstructed $\rho^-$ momentum, obtained from the vector sum of the measured $\pi^-$ and $\pi^0$ momenta and shown as a dashed blue line. For the $\pi\nu_\tau$ decay, the cone axis coincides with the measured $\pi^-$ momentum and is shown as a solid blue line. The dash-dotted black lines and shaded ellipse indicate the corresponding cone of kinematically allowed $\tau^-$ flight directions.}
\label{fig:tau_rho_decay_topology}
\end{figure}

The two signal channels considered in this study share a low-multiplicity hadronic topology with two charged pions and two undetected neutrinos. The $\pi\rho$ channel contains one $\pi^0\to\gamma\gamma$ decay and therefore two signal photons, whereas the $\rho\rho$ channel contains two $\pi^0$ mesons and four signal photons. The event reconstruction is performed separately for the two topologies and consists of charged-track and photon selection, $\pi^0$ and $\rho$ reconstruction~\cite{Sun_2025}, assignment of the visible particles to the two $\tau$ decay chains, multivariate background suppression, and reconstruction of the $\tau$ flight directions. Fig.~\ref{fig:tau_rho_decay_topology} illustrates the $\rho\nu_\tau$ decay topology and the kinematic ambiguity associated with the undetected neutrino.

\subsection{Charged-track selection}
\label{subsec:track_selection}

Charged tracks reconstructed in the MDC~\cite{achasov2024stcf,Ai:2023acts} are required to satisfy $|\cos\theta|<0.93$, where $\theta$ is the polar angle with respect to the beam axis. The distances of closest approach to the interaction point are required to be smaller than 20~cm along the beam direction and 10~cm in the transverse plane. Candidate events are required to contain exactly two good charged tracks with opposite electric charges. The same charged-track requirements are applied to the $\pi\rho$ and $\rho\rho$ channels.

Particle identification is performed using the standard global-likelihood estimator~\cite{achasov2024stcf,Qi:2021dtof}, which combines information from the specific ionization energy loss ($dE/dx$), time-of-flight system, and EMC. Both selected tracks are required to be consistent with the pion hypothesis. This requirement suppresses backgrounds containing electrons, muons, or kaons while retaining the charged pions from the signal $\tau$ decays.

\subsection{Photon selection}
\label{subsec:pi0_reco}

Photon candidates are reconstructed from EMC clusters~\cite{achasov2024stcf,Song:2023emc} with a minimum energy of 25~MeV in the barrel region ($|\cos\theta|<0.80$) and 50~MeV in the end-cap region ($0.86<|\cos\theta|<0.92$). EMC timing information is used to suppress out-of-time and beam-related backgrounds.

\begin{figure}[htbp]
    \centering 
        \includegraphics[width=\linewidth]{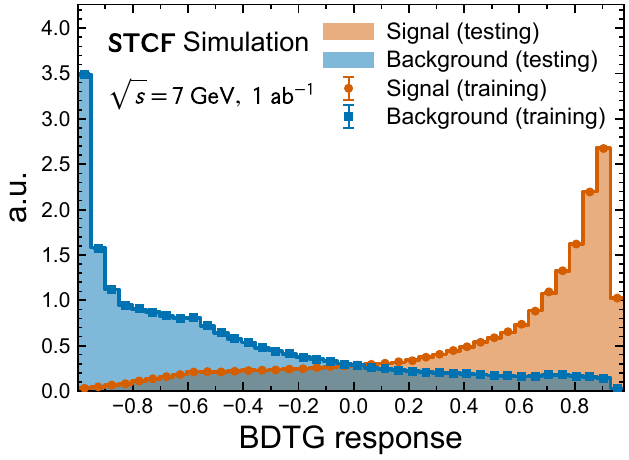}
        \caption{Normalized BDTG response distributions for signal photons from $\pi^0\to\gamma\gamma$ decays and ISR-related background clusters in the training and testing samples. The requirement $\mathrm{BDTG}>0$ is used for the photon-level selection.}
        \label{fig:ResponseGamma}
\end{figure}

To suppress soft and forward neutral clusters associated primarily with initial-state radiation (ISR), a dedicated gradient-boosted decision tree (BDTG) classifier~\cite{hoecker:2009} is applied at the photon-candidate level. The input observables are summarized in Table~\ref{tab:photon_bdt_inputs} and include the cluster energy, shower-shape variables, crystal multiplicity, and polar-angle information. The corresponding BDTG response distributions for signal photons from $\pi^0$ decays and ISR-related background clusters are shown in Fig.~\ref{fig:ResponseGamma}. Requiring $\mathrm{BDTG}>0$ retains approximately 94\% of signal photons while reducing the ISR-photon contribution by about one order of magnitude.

After the photon-level selection, $\pi\rho$ candidates are required to contain exactly two selected photons, while $\rho\rho$ candidates are required to contain exactly four. The selected photons are not assigned to individual $\pi^0$ candidates at this stage. Instead, all allowed photon pairings and their associations with the charged pions are considered simultaneously in the constrained kinematic-fit procedure described in Sec.~\ref{subsec:rho_reco}. This treatment accounts for the combinatorial ambiguity in the four-photon $\rho\rho$ topology and reduces contamination from incorrectly selected ISR-related clusters.

\begin{table}[tb]
\centering
\caption{EMC cluster observables used as inputs to the photon-selection BDTG classifier.}
\label{tab:photon_bdt_inputs}
\begin{tabular}{lll}
\hline
Observable & Definition & Description \\
\hline
$E_{\rm clus}$ & Cluster energy & Total energy deposited in the EMC cluster \\
$M_2$ & Second moment & Energy-weighted radial spread of the shower \\
$A_{42}$ & $a_{42}$ moment & Higher-order shower-shape observable \\
$N_{\rm hit}$ & Crystal multiplicity & Number of crystals associated with the cluster \\
$f_{3\times3}$ & Core-energy fraction & Energy concentration in the central $3\times3$ array \\
$R_{\rm ns}$ & Non-seed energy ratio & Mean non-seed energy normalized to the seed energy \\
$E_{\rm seed}$ & Seed energy & Energy deposited in the seed crystal \\
$\cos\theta_{\rm clus}$ & Cluster polar angle & Polar angle of the cluster direction \\
$L$ & Lateral moment & Observable characterizing the transverse shower profile \\
\hline
\end{tabular}
\end{table}

\subsection{$\pi^0$ and $\rho$ reconstruction with constrained kinematic fitting}
\label{subsec:rho_reco}

Following the charged-track and photon preselection, the visible particles are assigned to the two $\tau$ decay chains. The assignment simultaneously determines the photon pairing into one or two $\pi^0$ candidates and the association of the reconstructed $\pi^0$ meson(s) with the charged pions. The optimal configuration is selected using a constrained kinematic fit based on the Lagrange-multiplier method~\cite{YanLiang_2010}.

For the $\pi\rho$ topology, the two selected photons form a single $\pi^0$ candidate, which can be associated with either of the two charged pions, resulting in two assignment hypotheses per event. For the $\rho\rho$ topology, the four photons can be grouped into two $\pi^0$ candidates in three distinct ways, and each photon pairing can subsequently be associated with the two charged pions in two ways, giving six hypotheses per event.

For each hypothesis, the fit adjusts the measured charged-track and photon parameters together with the kinematic parameters associated with the two undetected neutrinos. Overall four-momentum conservation is imposed,
\begin{equation}
p_{e^+}+p_{e^-}
=
\sum_{\rm vis}p_i+p_{\nu_+}+p_{\nu_-},
\end{equation}
together with the appropriate $\pi^0$-mass constraint(s),
\begin{equation}
m_{\gamma\gamma}=m_{\pi^0},
\end{equation}
and the two $\tau$-mass constraints,
\begin{equation}
\left(p_{h^\pm}+p_{\nu_\pm}\right)^2=m_\tau^2,
\end{equation}
where $h^\pm$ denotes the visible hadronic system associated with the corresponding $\tau$ decay. For the $\pi\rho$ channel, $h^\pm$ is either a charged pion or a reconstructed $\rho$ candidate, whereas both visible systems $h^\pm$ correspond to reconstructed $\rho$ candidates in the $\rho\rho$ channel. One $\pi^0$-mass constraint is imposed for the $\pi\rho$ topology and two for the $\rho\rho$ topology.

The goodness of fit is quantified by
\begin{equation}
\chi^2_{\rm fit}
=
\left(\mathbf{x}_{\rm meas}-\mathbf{x}_{\rm fit}\right)^T
V^{-1}
\left(\mathbf{x}_{\rm meas}-\mathbf{x}_{\rm fit}\right),
\end{equation}
where $\mathbf{x}_{\rm meas}$ and $\mathbf{x}_{\rm fit}$ denote the measured and fitted detector observables, respectively, and $V$ is the corresponding measurement covariance matrix. The hypothesis with the smallest $\chi^2_{\rm fit}$ is retained.

\begin{figure}[htbp]
    \centering 
        \includegraphics[width=\linewidth]{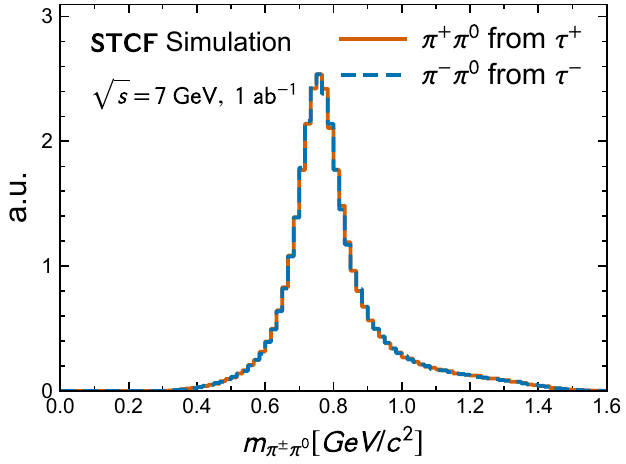}
        \caption{Normalized $\pi^\pm\pi^0$ invariant-mass distributions for the combinations assigned to the $\tau^+$ (orange) and $\tau^-$ (blue dashed) decay chains in the $\rho\rho$ signal sample after particle assignment and constrained kinematic fitting. The enhancement around the $\rho(770)$ mass demonstrates the reconstruction of the $\rho^\pm\to\pi^\pm\pi^0$ topology.} 
        \label{fig:pipi0Mass}
\end{figure}

\begin{figure}[htbp]
        \centering
        \includegraphics[width=\linewidth]{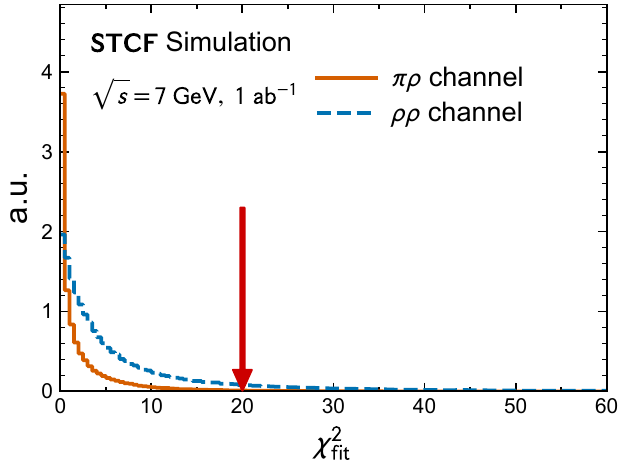}
        \caption{Distributions of the constrained kinematic-fit $\chi^2_{\rm fit}$ for the retained assignment hypotheses in the $\rho\rho$ (blue dashed) and $\pi\rho$ (orange) signal channels. The arrow indicates the requirement $\chi^2_{\rm fit}<20$ used to suppress incorrectly assigned and poorly reconstructed events.}
        \label{fig:ChisqDistrubution}
\end{figure}

For the $\rho\rho$ channel, the reconstructed $\pi^\pm\pi^0$ invariant-mass distributions after the assignment procedure are shown in Fig.~\ref{fig:pipi0Mass} and exhibit the expected enhancement around the $\rho(770)$ mass. Fig.~\ref{fig:ChisqDistrubution} compares the $\chi^2_{\rm fit}$ distributions of the selected hypotheses for the $\rho\rho$ and $\pi\rho$ channels. The high-$\chi^2_{\rm fit}$ tails are dominated by incorrect particle assignments and by events in which ISR-related clusters are incorporated into the reconstructed $\pi^0$ candidates. A common requirement of $\chi^2_{\rm fit}<20$ is therefore applied. In the signal simulation, truth matching shows that the fraction of correctly assigned events increases from 94.3\% to 97.9\% for the $\rho\rho$ channel and from 95.7\% to 99.0\% for the $\pi\rho$ channel. For the $\rho\rho$ channel, the residual misassignment rate involving ISR-related photons is reduced to approximately 2\%.

The fitting procedure is implemented within the STCF offline software framework using BESIII-derived kinematic-fitting utilities~\cite{YanLiang_2010}. Owing to the low visible-particle multiplicities of the two signal topologies, all assignment hypotheses can be tested explicitly on an event-by-event basis.

\subsection{Final multivariate selection}
\label{subsec:final_mva}

After the constrained kinematic fit, the dominant residual backgrounds arise from other $\tau^+\tau^-$ decay modes that share the selected charged-track and photon topology, together with non-resonant hadronic production. To further suppress these contributions, separate event-level BDTG classifiers are trained for the $\rho\rho$ and $\pi\rho$ channels after the particle-assignment procedure.

\begin{figure}[htbp]
    \centering 
        \centering
        \includegraphics[width=0.48\textwidth]{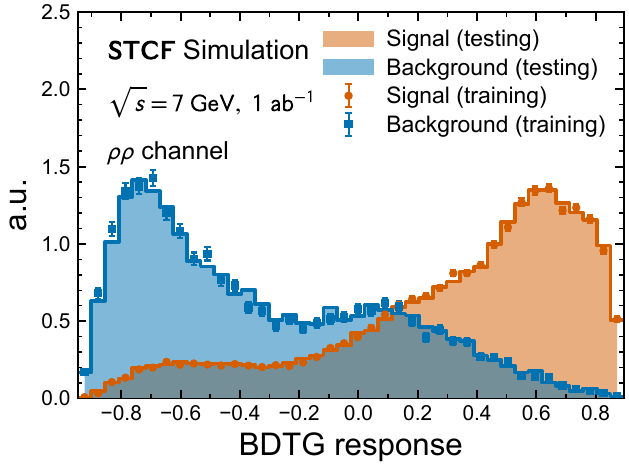}
        \hfill
        \includegraphics[width=0.48\textwidth]{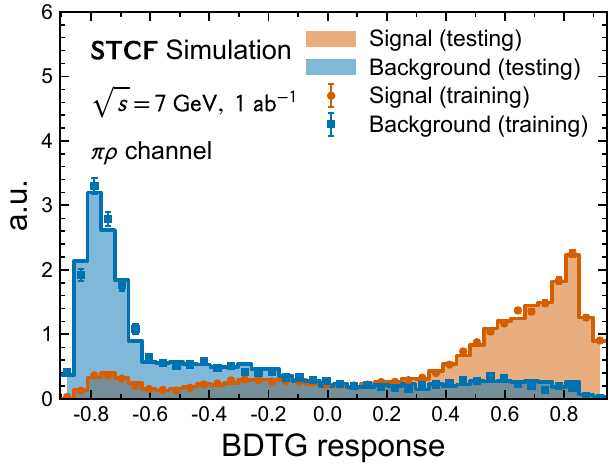}
        \caption{Event-level BDTG response distributions for the $\rho\rho$ (left) and $\pi\rho$ (right) channels. The filled distributions correspond to the testing samples, while the markers show the corresponding training samples for signal (orange) and background (blue).}
        \label{fig:ResponseEvent}
\end{figure}

The classifier inputs include the reconstructed masses and momenta of the visible hadronic systems, the missing-mass-squared, particle-identification information, visible-energy fractions, and additional kinematic variables sensitive to the signal topology. The BDTG response distributions for the $\rho\rho$ and $\pi\rho$ channels are shown in Fig.~\ref{fig:ResponseEvent}.

\begin{figure}[htbp]
    \centering 
    \begin{minipage}[t]{0.48\textwidth}
        \centering
        \includegraphics[width=\textwidth]{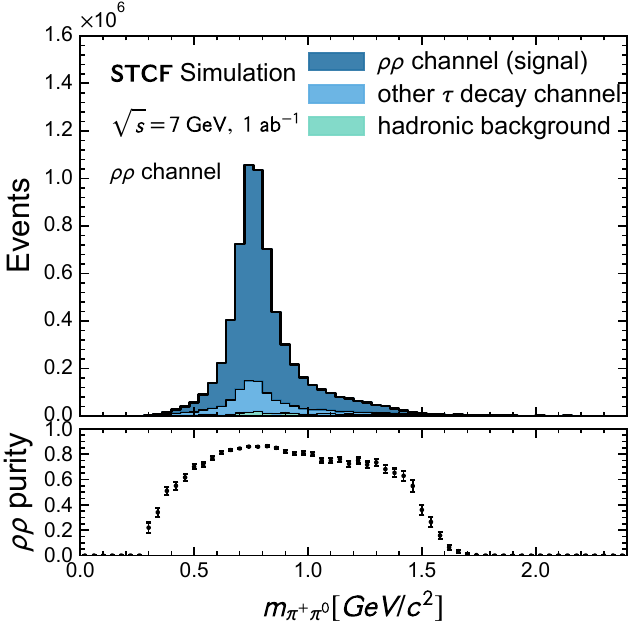}
        \caption{Invariant-mass distribution of the reconstructed $\pi^+\pi^0$ candidate in the $\rho\rho$ channel after the event-level BDTG selection. The upper panel shows the $\rho\rho$ signal, other $\tau$-decay backgrounds, and non-resonant hadronic background, while the lower panel shows the corresponding signal purity as a function of $m_{\pi^+\pi^0}$.}
        \label{fig:pippi0MassforSigBkg}
    \end{minipage}
    \hfill
    \begin{minipage}[t]{0.48\textwidth}
        \centering
        \includegraphics[width=\textwidth]{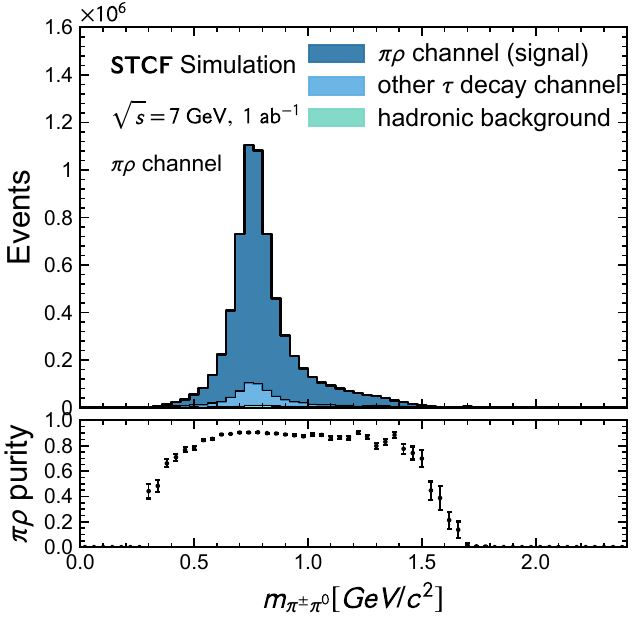}
        \caption{Invariant-mass distribution of the reconstructed $\pi^\pm\pi^0$ candidate in the $\pi\rho$ channel after the event-level BDTG selection. The upper panel shows the $\pi\rho$ signal, other $\tau$-decay backgrounds, and non-resonant hadronic background, while the lower panel shows the corresponding signal purity as a function of $m_{\pi^\pm\pi^0}$.}
        \label{fig:pipmpi0MassforSigBkg_pirho}
    \end{minipage}
\end{figure}

A common working point of $\mathrm{BDTG}>-0.2$ is adopted for the two channels, providing a favorable compromise between signal efficiency and residual-background rejection. The reconstructed $\pi^\pm\pi^0$ invariant-mass distributions after the multivariate selection are shown in Figs.~\ref{fig:pippi0MassforSigBkg} and~\ref{fig:pipmpi0MassforSigBkg_pirho} for the $\rho\rho$ and $\pi\rho$ channels, respectively. To further suppress non-resonant combinations, each reconstructed $\rho$ candidate is required to satisfy $0.6<m_{\pi^\pm\pi^0}<1.0~\mathrm{GeV}/c^2$; this requirement is applied to both $\rho$ candidates in the $\rho\rho$ channel and to the single $\rho$ candidate in the $\pi\rho$ channel.

\FloatBarrier

\subsection{$\tau$ flight-direction reconstruction}
\label{subsec:taurec}

Because both signal topologies contain two undetected neutrinos, the individual $\tau$ four-momenta cannot be reconstructed directly from the visible final state. The $\tau$ flight directions are nevertheless required to construct the decay polarimeters and the spin-density matrix. We therefore employ the standard two-cone kinematic construction, in which each one-prong hadronic $\tau$ decay constrains the parent $\tau$ direction to lie on a cone around the momentum of the corresponding visible hadronic system~\cite{gentile1996}, as illustrated in Fig.~\ref{fig:tau_rho_decay_topology}.

After reconstruction of the visible final state, the event is transformed to the $e^+e^-$ center-of-mass frame. Neglecting residual ISR effects, each $\tau$ lepton has energy $E_\tau=\sqrt{s}/2$ and the two $\tau$ leptons are produced back to back. For a decay $\tau^\pm\to h^\pm\nu_\tau$, where $h^\pm$ denotes either a charged pion or a reconstructed $\rho^\pm$, the opening angle between the $\tau$ direction and the visible hadronic momentum is determined by energy--momentum conservation:
\begin{equation}
\cos\theta_\pm=\frac{\gamma x_\pm-(1+r^2_\pm)/(2\gamma)}{\beta\sqrt{\gamma^2x^2_\pm-r^2_\pm}},
\end{equation}
where
\[
\begin{aligned}
x_\pm&=\frac{E_{h^\pm}}{E_\tau}, &
r_\pm&=\frac{m_{h^\pm}}{m_\tau}, &
\gamma&=\frac{E_\tau}{m_\tau}, &
\beta&=\sqrt{1-\frac{1}{\gamma^2}}.
\end{aligned}
\]
For a $\rho^\pm\nu_\tau$ decay, $h^\pm$ corresponds to the reconstructed $\pi^\pm\pi^0$ system, whereas for a $\pi^\pm\nu_\tau$ decay it corresponds to the charged pion alone. The intersection of the two kinematic cones yields, in general, two allowed solutions for the $\tau$ flight directions.

To reduce the twofold ambiguity, trajectory information from the charged pions is incorporated using the method described in Appendix~\ref{app:trajectory}. For each kinematic solution, the reconstructed charged-pion trajectories are compared with the corresponding production and decay geometry, and the solution with the smaller spatial inconsistency is selected. 

\begin{figure}[htbp]        
        \centering
        \includegraphics[width=0.48\textwidth]{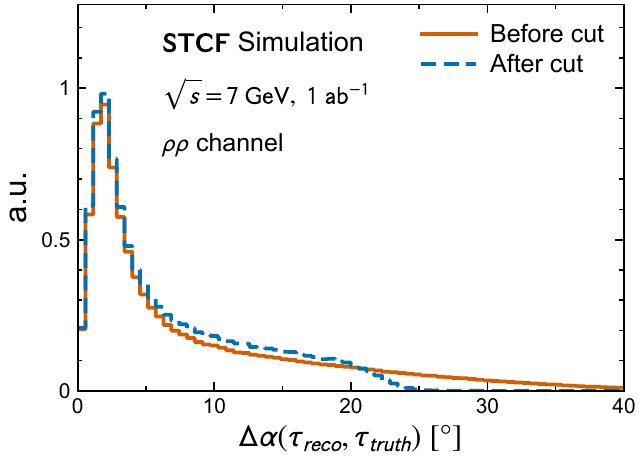}
        \hfill
        \includegraphics[width=0.48\textwidth]{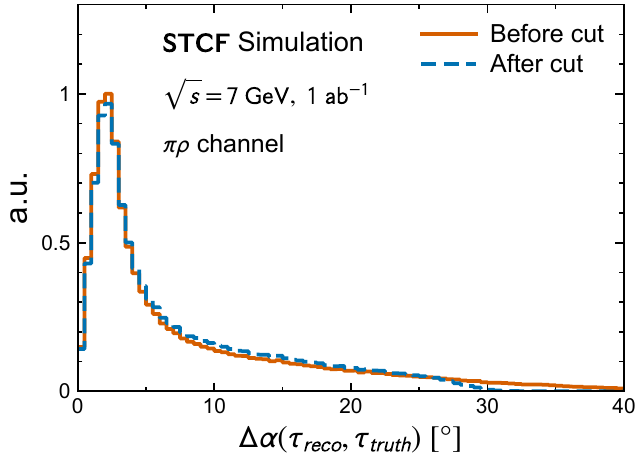}
        \caption{Angular difference $\Delta\alpha(\tau_{\rm reco},\tau_{\rm truth})$ between the reconstructed and truth $\tau$ directions before (orange) and after (blue dashed) the ambiguity-quality requirement $\sin(\alpha/2)<0.25$, for the $\rho\rho$ (left) and $\pi\rho$ (right) channels.}
        \label{fig:TauDirection}
\end{figure}

The residual impact of the twofold ambiguity is characterized by the opening angle $\alpha$ between the two candidate $\tau$ directions. Events for which the two solutions are widely separated are suppressed by requiring $\sin(\alpha/2)<0.25$. Fig.~\ref{fig:TauDirection} shows the angular difference $\Delta\alpha(\tau_{\rm reco},\tau_{\rm truth})$ before and after this ambiguity-quality requirement. The requirement reduces the RMS from approximately $159$~mrad to $120$~mrad ($122$~mrad) for the $\rho\rho$ ($\pi\rho$) channel, while the peak position remains at approximately $35$~mrad ($39$~mrad). 

\begin{figure}[htbp]        
        \centering
        \includegraphics[width=\linewidth]{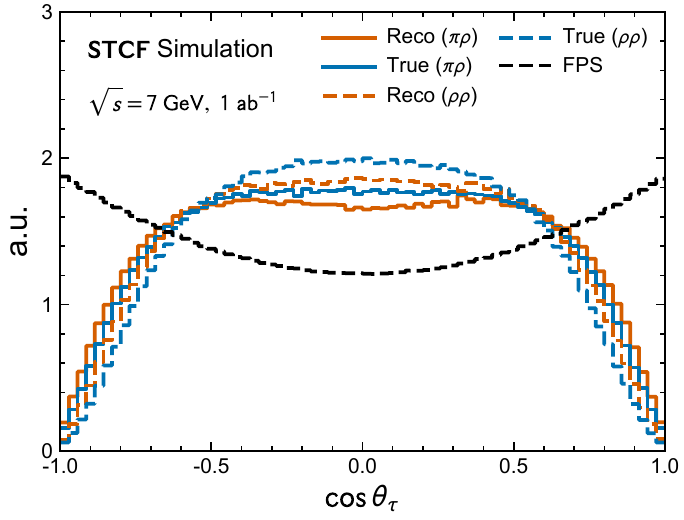}
        \caption{Reconstructed (orange) and truth-level (blue) \(\cos\theta_{\tau}\) distributions in the signal region, compared with \(\cos\theta_{\tau}\) in the full phase space (black), for the \(\rho\rho\) (dashed) and \(\pi\rho\) (solid) channels.}
        \label{fig:TauPolarAngle}
\end{figure}

Fig.~\ref{fig:TauPolarAngle} compares the reconstructed $\cos\theta_\tau$ distributions with the corresponding truth-level distributions in the selected signal samples, together with the truth-level distribution in the full phase space. Both channels exhibit a depletion toward large $|\cos\theta_\tau|$, reflecting the reduced detector acceptance and reconstruction performance in the forward region. This behavior motivates the use of a central fiducial region in the extraction of the spin-correlation and entanglement-sensitive observables discussed in Sec.~\ref{sec:result}.

\FloatBarrier

\subsection{Selection performance and expected yield}
\label{subsec:selection_performance}

The cumulative performance of the event selection is summarized in Table~\ref{tab:cutflow} for the $\rho\rho$ and $\pi\rho$ channels. Relative to the corresponding generated signal samples, the overall selection efficiencies are approximately 4.5\% and 4.8\%, respectively. When the signal branching fractions are included, the selected fractions relative to inclusive $e^+e^-\to\tau^+\tau^-$ production are approximately 0.43\% for $\rho\rho$ and 0.26\% for $\pi\rho$. At the same time, the signal purity increases from 1.3\% to 87.6\% for the $\rho\rho$ channel and from 1.2\% to 93.2\% for the $\pi\rho$ channel.

Under the nominal STCF annual-running scenario adopted in this study, approximately $1.9\times10^9$ $\tau^+\tau^-$ pairs are expected at $\sqrt{s}=7$~GeV. After accounting for the relevant branching fractions, this corresponds to approximately $1.19\times10^8$ produced $\rho\rho$ events and $0.95\times10^8$ produced $\pi\rho$ events. Applying the full reconstruction and selection chain yields approximately $5.4\times10^6$ and $4.6\times10^6$ selected signal events, respectively. These event samples provide the statistical basis for the detector-level spin-correlation study presented in Sec.~\ref{sec:result}.

\begin{table}[!t]
    \centering
    \scriptsize
    \caption{Cut-flow summary for the $\rho\rho$ (top) and $\pi\rho$ (bottom) signal channels and the dominant background categories. Except for the final cumulative row, the quoted signal efficiencies and background surviving fractions are evaluated relative to the preceding selection step. The final row gives the cumulative efficiency relative to the generated sample. The signal purity after each selection stage is given in the last column.}
    \label{tab:cutflow}
    \begin{tabular}{p{5.7cm}cccc}
\hline
Selection step & $\rho\rho$ signal eff. (\%) & \makecell[c]{Non-resonant hadronic\\ bkg. (\%)} & \makecell[c]{Other $\tau^+\tau^-$\\ bkg. (\%)} & \makecell[c]{Signal purity\\ (\%)} \\
\hline
Generated sample & -- & -- & -- & 1.3 \\
Two good charged tracks with opposite charge sign & 77.8 & 6.5 & 56.2 & 5.5 \\
Exactly four photon candidates & 21.8 & 15.6 & 5.8 & 13.0 \\
Two identified pions ($N_{\pi^+}=1$, $N_{\pi^-}=1$) & 79.6 & 51.8 & 59.9 & 17.6 \\
Accepted pairing and kinematic-fit preselection & 45.3 & 2.7 & 5.4 & 70.1 \\
Event-level BDTG requirement & 87.8 & 51.3 & 37.6 & 83.1 \\
$\tau$-direction reconstruction requirement & 99.4 & 61.6 & 82.5 & 86.8 \\
Final mass and reconstruction-quality requirements & 83.4 & 71.7 & 50.9 & 87.6 \\
\hline
Overall cumulative efficiency & 4.5 & $<0.01$ & $<0.03$ & -- \\
\hline
\end{tabular}
\vspace{0.4cm}
\begin{tabular}{p{5.7cm}cccc}
\hline
Selection step ($\pi\rho$ channel) & $\pi\rho$ signal eff. (\%) & \makecell[c]{Non-resonant hadronic\\ bkg. (\%)} & \makecell[c]{Other $\tau^+\tau^-$\\ bkg. (\%)} & \makecell[c]{Signal purity\\ (\%)} \\
\hline
Generated sample & -- & -- & -- & 1.2 \\
Two good charged tracks with opposite charge sign & 52.1 & 6.5 & 56.8 & 3.4 \\
Exactly two photon candidates & 43.2 & 15.0 & 21.5 & 7.2 \\
Two identified pions ($N_{\pi^+}=1$, $N_{\pi^-}=1$) & 73.7 & 41.3 & 23.1 & 17.5 \\
Accepted pairing and kinematic-fit preselection & 43.5 & 1.3 & 3.0 & 79.1 \\
Event-level BDTG requirement & 87.0 & 65.9 & 28.7 & 90.3 \\
$\tau$-direction reconstruction requirement & 96.6 & 73.1 & 94.7 & 91.1 \\
Final mass and reconstruction-quality requirements & 79.0 & 20.4 & 74.1 & 93.2 \\
\hline
Overall cumulative efficiency & 4.8 & $<0.01$ & $<0.03$ & -- \\
\hline
\end{tabular}
\end{table}

\FloatBarrier

\section{Entanglement-sensitive observables}
\label{sec:result}

\subsection{Spin-density-matrix reconstruction}
\label{sec:sdm_reco}

Following the formalism introduced in Sec.~\ref{sec:theory}, the $\tau^+\tau^-$ spin-density matrix is reconstructed separately for the $\rho\rho$ and $\pi\rho$ channels from the polarimetric information carried by the visible $\tau$ decay products. After the event reconstruction and $\tau$-direction determination described in Sec.~\ref{sec:selection}, the decay products are boosted to the corresponding reconstructed $\tau^\pm$ rest frames. The polarimetric vectors $\vec h^{\,\pm}$ are then constructed event by event using Eq.~\eqref{eq:h_pi_mode} for a $\pi\nu_\tau$ decay and Eq.~\eqref{eq:rho_polarimeter} for a $\rho\nu_\tau$ decay. The polarization coefficients $B_i^\pm$ and spin-correlation coefficients $C_{ij}$ are obtained from the corresponding sample averages according to Eqs.~(\ref{eq:Bi_def}) and~(\ref{eq:Cij_def}). All polarization components $B_i^\pm$ and all nine elements $C_{ij}$ are reconstructed and used in the evaluation of the spin-density matrix and the derived observables. 

\begin{figure}[htbp]
  \centering
  \includegraphics[width=0.98\textwidth]{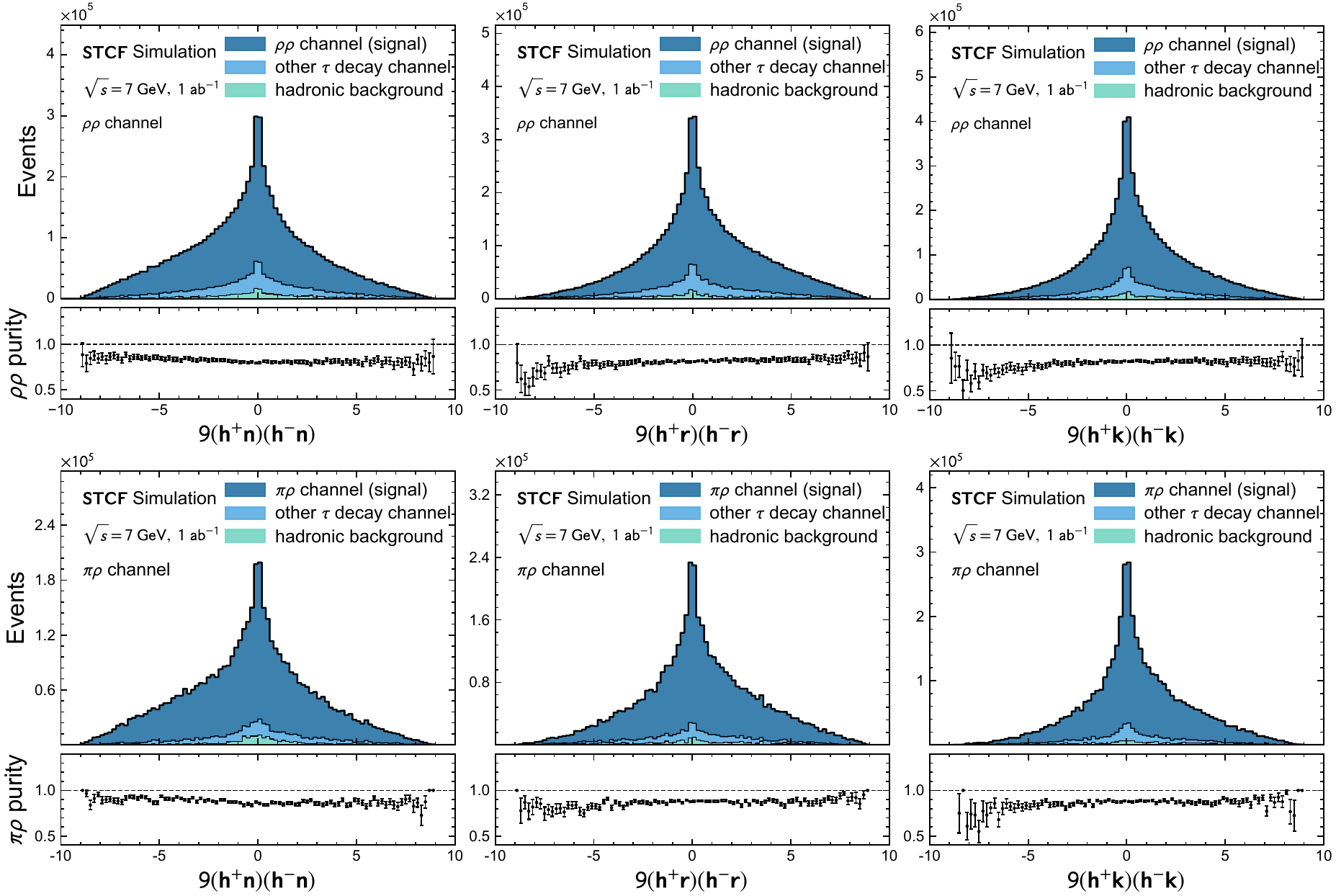}
  \vspace{0.1cm}
    \caption{Distributions of the three diagonal bilinear polarimetric combinations $h_n^+h_n^-$, $h_r^+h_r^-$, and $h_k^+h_k^-$, corresponding to $C_{11}$, $C_{22}$, and $C_{33}$, for the selected $\rho\rho$ and $\pi\rho$ channels. In each row, the upper panels show the corresponding signal contribution, other $\tau$-decay backgrounds, and non-resonant hadronic background, while the lower panels show the signal purity.}
  \label{fig:SDMforSigBkg}
\end{figure}

Fig.~\ref{fig:SDMforSigBkg} shows the three diagonal bilinear polarimetric combinations $h_n^+h_n^-$, $h_r^+h_r^-$, and $h_k^+h_k^-$, corresponding to $C_{11}$, $C_{22}$, and $C_{33}$, for the selected $\rho\rho$ and $\pi\rho$ samples. These representative components provide a direct illustration of the polarimetric information entering the reconstructed correlation matrix. The selected samples are dominated by the corresponding signal contributions, with smaller residual components from other $\tau$ decays and non-resonant hadronic production. In the subsequent extraction, the reconstructed observables are evaluated for the selected signal-plus-background samples using the simulated background composition.

\begin{figure}[htbp]
  \centering
  \includegraphics[width=0.98\textwidth]{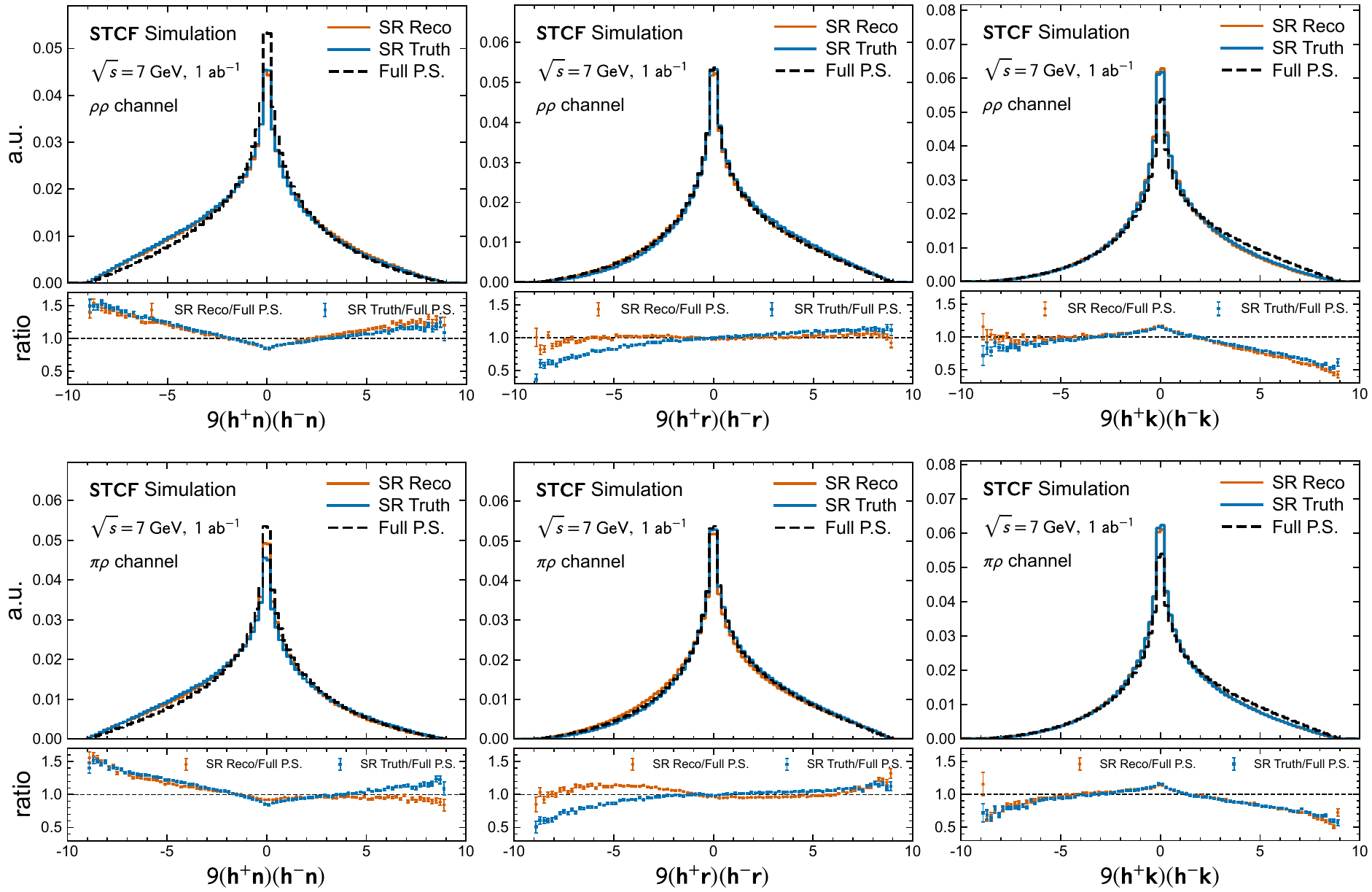}
    \caption{Comparison of the three diagonal bilinear polarimetric combinations $h_n^+h_n^-$, $h_r^+h_r^-$, and $h_k^+h_k^-$ for the $\rho\rho$ and $\pi\rho$ channels. The upper panels show the reconstructed distributions in the selected signal region (SR Reco), the corresponding truth-level distributions in the selected region (SR Truth), and the truth-level distributions in the full phase space (Full P.S.). The lower panels show the ratios of the selected-region reconstructed and truth-level distributions to the corresponding full-phase-space truth distributions.}
  \label{fig:SDMforSR&FPS}
\end{figure}

Fig.~\ref{fig:SDMforSR&FPS} compares the reconstructed distributions of the same three diagonal combinations with their truth-level counterparts in the selected signal region and with the corresponding truth-level distributions in the full phase space. The difference between the selected-region truth and full-phase-space distributions reflects the effect of the event selection and detector acceptance, whereas the additional difference between reconstructed and selected-region truth distributions quantifies reconstruction-induced distortions. These effects are most visible toward the forward region, where the acceptance and $\tau$-direction reconstruction deteriorate, as already indicated by Fig.~\ref{fig:TauPolarAngle}.

These acceptance and reconstruction effects motivate evaluating the final observables in a restricted fiducial region. The nominal requirement $|\cos\theta_\tau|<0.2$ is chosen to emphasize the central production region, where the intrinsic spin-correlation sensitivity is enhanced and the detector response is comparatively uniform. Accordingly, the quantities reported below follow the detector-level fiducial interpretation introduced in Sec.~\ref{subsec:sdm}; no unfolding or extrapolation to the full angular phase space is performed.

From the reconstructed SDM, the concurrence $C[\rho]$ and the Bell-sensitive quantity $m_{12}[C]$ are evaluated as defined in Sec.~\ref{subsec:entangle_sensitive}. The concurrence is used to characterize entanglement in the reconstructed two-spin state, while $m_{12}[C]$ is interpreted as a Bell-sensitive observable derived from the reconstructed spin-correlation matrix rather than as the outcome of a loophole-free Bell test.

\subsection{Statistical and detector-resolution uncertainties}
\label{sec:uncertainties}

Statistical uncertainties on the reconstructed SDM coefficients and on the derived observables $C[\rho]$ and $m_{12}[C]$ are estimated using bootstrap resampling~\cite{efron1992bootstrap}. For each channel, $N_{\rm toy}=1000$ bootstrap replicas are generated from the final selected sample by sampling events with replacement while preserving the nominal sample size, and the complete observable-extraction procedure is repeated for each replica. For a generic observable $X$, the statistical uncertainty is defined as $\delta X_{\rm stat}=(X_{84}-X_{16})/2$, where $X_{16}$ and $X_{84}$ denote the 16th and 84th percentiles of the bootstrap distribution. The resulting uncertainties are determined from the available simulated samples and subsequently scaled to the benchmark integrated luminosity of $1~\mathrm{ab}^{-1}$ according to $\delta X_{\rm stat}\propto L^{-1/2}$, assuming that the selected event composition and reconstruction performance remain unchanged with luminosity. The quoted statistical uncertainties therefore represent the projected statistical precision rather than the raw uncertainty of the finite Monte Carlo samples.

Among the experimental effects that can influence the reconstructed spin-correlation observables, the contribution associated with finite detector resolution is evaluated explicitly, as it propagates directly through the complete analysis chain and is the contribution most relevant to detector optimization. It is estimated with a covariance-based smearing procedure applied to the simulated signal samples. Charged-track parameters are varied according to the covariance matrices provided by the tracking reconstruction, while photon energies and directions are varied using the corresponding EMC covariance information. The correlated variations are propagated through the complete analysis chain, including particle assignment, constrained kinematic fitting, $\tau$-direction reconstruction, SDM extraction, and evaluation of the derived observables. An ensemble of smeared replicas is generated, and the reconstructed observables are recalculated for each realization. The detector-resolution uncertainty $\delta X_{\rm det}$ is defined analogously from the 16th and 84th percentiles of the smeared distribution. Further details of the procedure are given in Appendix~\ref{app:detector_uncertainty}.

As the present work is a feasibility study aimed at validating the analysis method and guiding detector optimization, uncertainty sources that can only be derived based on real data, such as reconstruction efficiencies and acceptance corrections, are beyond its scope. A recent BESIII measurement found such efficiency-correction uncertainties to be small relative to the statistical precision~\cite{BESIII:2026LambdaCP}, supporting their neglect in the present study. Theoretical uncertainties from the signal and residual-background modelings are also not considered. The quoted sensitivity is consequently a detector-level projection combining statistical and detector-resolution effects, rather than a complete forecast for a future measurement.

\subsection{Fiducial optimization and projected sensitivity}
\label{subsec:sensitivity}

\begin{figure}[!ht]
        \centering
        \includegraphics[width=0.48\textwidth]{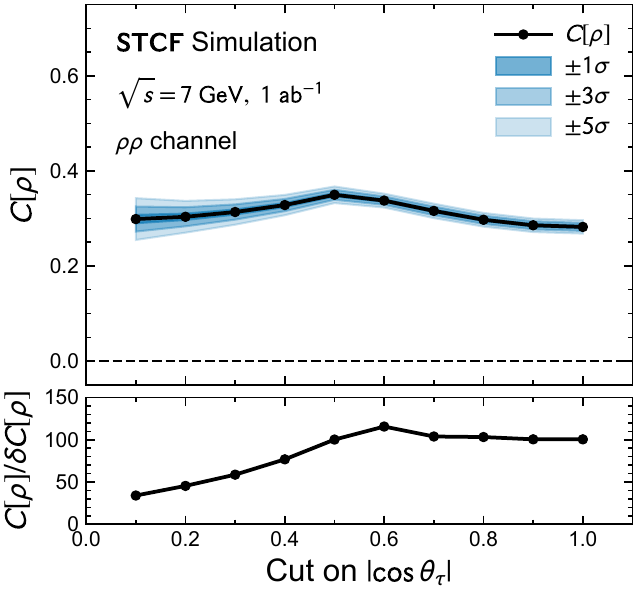}
        \hfill
        \includegraphics[width=0.48\textwidth]{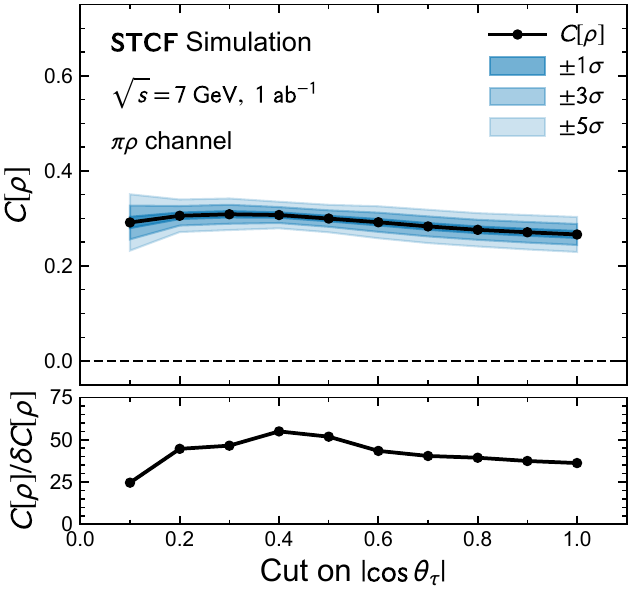}
        \caption{Concurrence $C[\rho]$ (top) and projected significance $Z_C=C[\rho]/\delta C_{\rm tot}$ (bottom) as functions of the fiducial upper bound on $|\cos\theta_\tau|$, for the $\rho\rho$ (left) and $\pi\rho$ (right) channels. The shaded regions indicate the $\pm1\sigma$, $\pm3\sigma$, and $\pm5\sigma$ intervals obtained from the combined statistical and detector-resolution uncertainty $\delta C_{\rm tot}$.}
        \label{fig:result1}
\end{figure}

\begin{figure}[!ht]
        \centering
        \includegraphics[width=0.48\textwidth]{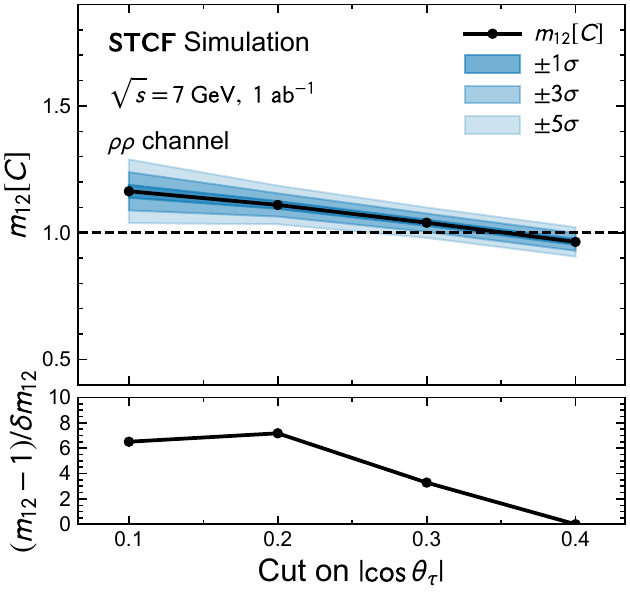}
        \hfill
        \includegraphics[width=0.48\textwidth]{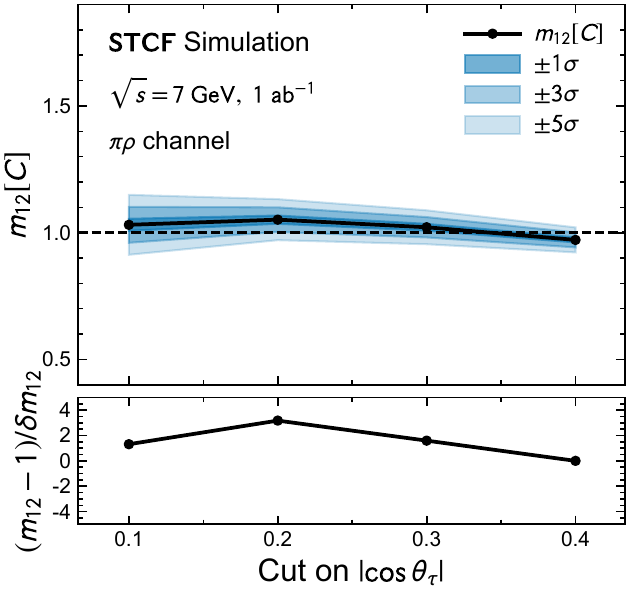}
        \caption{Bell-sensitive quantity $m_{12}[C]$ (top) and projected significance $Z_{m_{12}}=(m_{12}[C]-1)/\delta m_{12,\rm tot}$ relative to the reference threshold $m_{12}[C]=1$ (bottom) as functions of the fiducial upper bound on $|\cos\theta_\tau|$, for the $\rho\rho$ (left) and $\pi\rho$ (right) channels. The shaded regions indicate the corresponding $\pm1\sigma$, $\pm3\sigma$, and $\pm5\sigma$ intervals based on the combined statistical and detector-resolution uncertainty.}
        \label{fig:result2}
\end{figure}

The dependence of the reconstructed entanglement-sensitive observables on the fiducial acceptance is studied by varying the maximum allowed value of $|\cos\theta_\tau|$, where $\theta_\tau$ denotes the reconstructed $\tau$ production angle in the $\tau^+\tau^-$ center-of-mass frame. Figures~\ref{fig:result1} and~\ref{fig:result2} show the corresponding behavior of $C[\rho]$ and $m_{12}[C]$ for the $\rho\rho$ and $\pi\rho$ channels. The scan illustrates the balance between the increasing intrinsic sensitivity toward central production and the loss of event yield associated with a tighter angular requirement. For the benchmark sensitivity estimate, the statistical and detector-resolution contributions are treated as independent and combined in quadrature,
\[
\delta X_{\rm tot}=\sqrt{\delta X_{\rm stat}^2+\delta X_{\rm det}^2}.
\]
The significance for the concurrence is defined as
\[
Z_C=\frac{C[\rho]}{\delta C_{\rm tot}},
\]
while that for the Bell-sensitive observable is evaluated relative to the reference threshold $m_{12}[C]=1$,
\[
Z_{m_{12}}=\frac{m_{12}[C]-1}{\delta m_{12,{\rm tot}}}.
\]

Fig.~\ref{fig:result1} shows the concurrence and its significance relative to the separable-state reference value $C[\rho]=0$. For both channels, the reconstructed concurrence remains positive throughout the fiducial range considered. The $\rho\rho$ channel provides the larger projected significance, while the $\pi\rho$ channel exhibits a similar central value but a larger total uncertainty. The dependence on the angular requirement is moderate compared with that observed for $m_{12}[C]$, indicating that the concurrence is relatively stable against variations of the fiducial acceptance.

Fig.~\ref{fig:result2} shows the corresponding results for $m_{12}[C]$. The observable exhibits a stronger dependence on the fiducial requirement than the concurrence. In the $\rho\rho$ channel, restricting the sample to the central region increases the separation from the reference value $m_{12}[C]=1$ and yields the highest projected sensitivity near the nominal working point. The $\pi\rho$ channel shows a substantially weaker separation from the reference threshold because of its larger reconstructed uncertainty, despite having a comparable overall selected event yield. This difference reflects the channel dependence of the polarimetric reconstruction and should be interpreted together with the statistical and detector-resolution contributions discussed above.

For the nominal requirement $|\cos\theta_\tau|<0.2$ and an integrated luminosity of $1~\mathrm{ab}^{-1}$, the benchmark results are summarized in Table~\ref{tab:benchmark_results}. The concurrence is resolved with high significance in both channels, with the $\rho\rho$ channel providing the stronger sensitivity. For $m_{12}[C]$, the $\rho\rho$ channel exceeds the reference threshold $m_{12}[C]=1$ with a projected significance above $5\sigma$, whereas the $\pi\rho$ channel provides substantially lower sensitivity. The $\rho\rho$ topology therefore constitutes the primary channel for the Bell-sensitive observable in the present reconstruction strategy, while the $\pi\rho$ channel provides an additional and experimentally distinct probe of the reconstructed spin correlations.

\begin{table}[htbp]
    \centering
    \small
    \caption{Benchmark detector-level results in the nominal fiducial region $|\cos\theta_\tau|<0.2$ for the $\rho\rho$ and $\pi\rho$ channels, assuming an integrated luminosity of $1~\mathrm{ab}^{-1}$. The statistical and detector-resolution uncertainties are combined in quadrature to obtain $\delta_{\rm tot}$. For $C[\rho]$, the significance is evaluated relative to $C[\rho]=0$; for $m_{12}[C]$, it is evaluated relative to the reference threshold $m_{12}[C]=1$.}
    \label{tab:benchmark_results}
    \begin{tabular}{llccccc}
\hline
Channel & Observable & Central value & $\delta_{\rm stat}$ & $\delta_{\rm det}$ & $\delta_{\rm tot}$ & Significance \\
\hline
\multirow{2}{*}{$\rho\rho$} & $C[\rho]$ & $0.303$ & $0.003$ & $0.006$ & $0.007$ & $45.4$ \\
  & $m_{12}[C]$ & $1.109$ & $0.006$ & $0.014$ & $0.015$ & $7.2$ \\
\hline
\multirow{2}{*}{$\pi\rho$} & $C[\rho]$ & $0.306$ & $0.003$ & $0.006$ & $0.007$ & $44.6$ \\
  & $m_{12}[C]$ & $1.051$ & $0.007$ & $0.015$ & $0.016$ & $3.2$ \\
\hline
    \end{tabular}
\end{table}

The two channels exhibit comparable reconstructed concurrence values but different sensitivities once statistical and detector-resolution effects are included. The $\rho\rho$ channel provides the strongest sensitivity to both observables and, in particular, to the Bell-sensitive quantity $m_{12}[C]$, while the $\pi\rho$ channel offers an independent topology with higher selected purity but larger uncertainties in the reconstructed spin observables. Taken together, the two channels demonstrate the feasibility of extracting non-classical spin correlations from realistic hadronic $\tau$ decays at STCF and provide a basis for future combined analyses incorporating additional decay modes and a more complete treatment of experimental systematics.

\FloatBarrier

\section{Conclusions}
\label{sec:conclusion}

We have presented a detector-level feasibility study of entanglement-sensitive observables in $e^+e^-\to\tau^+\tau^-$ at STCF at $\sqrt{s}=7$~GeV, using the $\rho\rho$ and $\pi\rho$ final states. Signal and dominant background processes are simulated with the full STCF detector and reconstruction framework, and dedicated selections are developed for the two topologies. The resulting overall signal efficiencies are approximately $4.5\%$ for the $\rho\rho$ channel and $4.8\%$ for the $\pi\rho$ channel, with corresponding signal purities of about $87.6\%$ and $93.2\%$, respectively.

From the reconstructed decay kinematics, the $\tau^+\tau^-$ spin-density matrix is inferred and the concurrence $C[\rho]$ and Bell-sensitive quantity $m_{12}[C]$ are evaluated in a selected fiducial region. For the benchmark requirement $|\cos\theta_\tau|<0.2$ and an integrated luminosity of $1~\mathrm{ab}^{-1}$, the reconstructed concurrence is determined with a projected relative precision of about $2.2\%$ in both channels when statistical and detector-resolution uncertainties are combined. For the Bell-sensitive observable, the $\rho\rho$ channel gives $m_{12}[C]=1.1094$ and exceeds the reference threshold $m_{12}[C]=1$ with a projected significance of about $7.2\sigma$, whereas the $\pi\rho$ channel exhibits substantially lower sensitivity with a projected significance of about $3.2\sigma$ to provide an experimentally complementary decay topology.

These results demonstrate that non-classical spin correlations in $\tau^+\tau^-$ production can be reconstructed with realistic detector response in hadronic $\tau$ decays at STCF. The $\rho\rho$ topology provides the strongest sensitivity to both observables in the present analysis, while the $\pi\rho$ channel offers an independent high-purity sample and complementary polarimetric information. The quoted sensitivities incorporate statistical and detector-resolution effects and should therefore be regarded as detector-level projections rather than complete experimental forecasts. Future studies should extend the treatment of systematic uncertainties, improve acceptance corrections, and combine additional $\tau$ decay modes to further enhance the sensitivity to quantum correlations.



\backmatter

\subsection*{Funding}
The authors thank the members of the STCF software and physics groups for helpful discussions and technical support. This work was supported by the National Natural Science Foundation of China (Grants No.~12305119 and No.~12505116), the National Science and Technology Major Project (Grants No. 2026ZD0126000 and No. 2026ZD0126005), and the China Postdoctoral Science Foundation (Grant No.~2024M763123).

\subsection*{Data availability statement}
The datasets generated and/or analysed during the current study are available from the corresponding authors on reasonable request.

\subsection*{Code availability statement}
The analysis code used in this study is available from the corresponding authors on reasonable request.

\subsection*{Author contributions}
C.~Bao and X.~Tao performed the simulation and analysis and prepared the first draft of the manuscript. H.~Chen, M.~Liu, L.~Xu, and X.~Zhou contributed to the analysis design, interpretation of the results, and revision of the manuscript. All authors reviewed and approved the final manuscript.

\begin{appendices}

\section{Resolution of the $\tau$-direction ambiguity using trajectory information}
\label{app:trajectory}

For a one-prong hadronic decay $\tau^\pm\to h^\pm\nu_\tau$, with $h^\pm=\pi^\pm$ or $\rho^\pm$, the kinematic constraints described in Sec.~4.5 restrict the parent $\tau$ direction to a cone around the visible hadronic momentum. The intersection of the two cones associated with the $\tau^+$ and $\tau^-$ decays generally gives two kinematically allowed solutions. If unresolved, this twofold ambiguity degrades the reconstructed $\tau$ rest frames and consequently dilutes the spin-sensitive angular information.

To discriminate between the two solutions, trajectory information from the charged pions is incorporated following the strategy of Ref.~\cite{PhysRevD.107.093002}. The finite lifetime of the $\tau$ lepton, $c\tau\simeq87~\mu\mathrm{m}$, leads to a displaced decay point relative to the interaction region. The measured charged-track trajectory, in particular its impact-parameter information, therefore provides an additional geometrical constraint on the parent $\tau$ flight direction.

For each kinematically allowed solution, the expected charged-track geometry is evaluated under the corresponding $\tau$-direction hypothesis and compared with the measured track parameters. A likelihood is constructed in the generic form
\begin{equation}
\begin{aligned}
\ln\mathcal{L}(\mathbf{p})={}&-\frac{1}{2}\sum_i
\left[\frac{O_i-P_i(\mathbf{p})}{\sigma_i}\right]^2+\ln P_{\rm kin}(\mathbf{p}).
\end{aligned}
\label{eq:trajectory_likelihood}
\end{equation}
where $\mathbf{p}$ denotes the parameters specifying a candidate kinematic solution, $O_i$ are the measured trajectory observables, $P_i(\mathbf{p})$ are their predicted values for that solution, $\sigma_i$ are the corresponding resolutions, and $P_{\rm kin}$ incorporates the kinematic constraints of the event.

The preferred $\tau$-direction solution is selected according to the trajectory consistency encoded in Eq.~\eqref{eq:trajectory_likelihood}. The physically correct solution is expected to provide better agreement between the measured charged-track geometry and that predicted from the hypothesized $\tau$ flight directions. As discussed in Sec.~\ref{subsec:taurec}, the use of this information reduces the RMS of the opening angle $\Delta\alpha(\tau_{\rm reco},\tau_{\rm truth})$ from approximately $169$~mrad to $159$~mrad for the $\rho\rho$ channel.

Residual ambiguity is characterized by the opening angle $\alpha$ between the two candidate $\tau$ directions. Events with poorly separated solutions are retained preferentially through the requirement $\sin(\alpha/2)<0.25$, which reduces the impact of the remaining ambiguity on the reconstructed rest-frame observables.

\section{Propagation of detector-resolution uncertainties}
\label{app:detector_uncertainty}

The effect of finite detector resolution on the reconstructed spin observables is evaluated by propagating the covariance information associated with the reconstructed charged tracks and EMC clusters through the complete analysis chain. The procedure is applied to the signal Monte Carlo samples reconstructed with the full Geant4-based STCF detector simulation and the OSCAR offline framework~\cite{huang:2023}.

For each charged track, the reconstruction provides a covariance matrix describing the correlated uncertainties of the fitted track parameters, including the reconstructed position and momentum components. For photon candidates, the corresponding EMC covariance information describes the uncertainties and correlations associated with the reconstructed cluster energy and direction. These covariance matrices are used to generate correlated variations of the reconstructed particle parameters.

For a vector of reconstructed parameters $\mathbf{x}$ with covariance matrix $\Sigma$, a varied realization is constructed as
\begin{equation}
\mathbf{x}_{\rm smear}
=
\mathbf{x}_{\rm reco}
+
\delta\mathbf{x},
\qquad
\delta\mathbf{x}\sim\mathcal{N}(0,\Sigma).
\label{eq:covariance_smearing}
\end{equation}
Equivalently, the correlated variation may be generated as
\begin{equation}
\delta\mathbf{x}=A\mathbf{z},
\qquad
AA^{T}=\Sigma,
\label{eq:covariance_factorization}
\end{equation}
where the components of $\mathbf{z}$ are independent standard-normal random variables. This construction preserves the correlations among the reconstructed quantities encoded in the detector covariance matrix.

For each smeared realization, the affected particle four-momenta are recomputed and propagated through the subsequent reconstruction and analysis steps, including particle assignment, constrained kinematic fitting, $\tau$-direction reconstruction, spin-density-matrix extraction, and calculation of $C[\rho]$ and $m_{12}[C]$. An ensemble of such realizations is generated independently for the $\rho\rho$ and $\pi\rho$ signal samples.

For a reconstructed quantity $X$, the detector-resolution uncertainty is defined from the central 68\% interval of the resulting distribution,
\begin{equation}
\delta X_{\rm det}
=
\frac{X_{84}-X_{16}}{2},
\label{eq:detector_uncertainty}
\end{equation}
where $X_{16}$ and $X_{84}$ denote the 16th and 84th percentiles, respectively. The use of the full covariance information preserves correlations among detector observables that would be neglected by independent one-dimensional variations and is particularly relevant for quantities entering the constrained kinematic fit and the reconstruction of angular observables. A dedicated high-statistics signal sample is used to reduce fluctuations in this estimate. 


As emphasized in Sec.~\ref{sec:uncertainties}, $\delta X_{\rm det}$ represents the detector-resolution contribution considered in the present feasibility study and should not be interpreted as the complete experimental systematic uncertainty. Other effects, including reconstruction-efficiency uncertainties, residual-background modeling, generator dependence, and acceptance corrections, require dedicated evaluations in a future measurement.

\end{appendices}

\bibliographystyle{bst/spphys}
\bibliography{refs}

\end{document}